\documentclass[11pt,a4paper]{article}

\usepackage{graphicx}
\usepackage{afterpage}
\usepackage{epsfig,cite}
\usepackage{amssymb}
\usepackage{amsmath}
\usepackage{dsfont}
\usepackage{multirow}
\usepackage{url,hyperref}
\usepackage{bm}

\usepackage{url}
\usepackage{hyperref}

\usepackage{tikz}
\usetikzlibrary{shapes,arrows}

\newcommand{\be}{\begin{equation}}
\newcommand{\ee}{\end{equation}}
\newcommand{\bea}{\begin{eqnarray}}
\newcommand{\eea}{\end{eqnarray}}
\newcommand{\bi}{\begin{itemize}}
\newcommand{\ei}{\end{itemize}}
\newcommand{\ben}{\begin{enumerate}}
\newcommand{\een}{\end{enumerate}}

\newcommand{\lc}{\left[}
\newcommand{\rc}{\right]}
\newcommand{\lp}{\left(}
\newcommand{\rp}{\right)}

\def\frac#1#2{{{#1}\over {#2}}}
\def\gsim{\mathrel{\rlap{\lower4pt\hbox{\hskip1pt$\sim$}}
    \raise1pt\hbox{$>$}}}         
\def\lsim{\mathrel{\rlap{\lower4pt\hbox{\hskip1pt$\sim$}}
    \raise1pt\hbox{$<$}}}         

\newcommand{\draft}[1]{}

\def\beq{\begin{equation}}  
\def\eeq{\end{equation}}  
\usepackage{xspace}

\def \n0{N_j^{(0)}}

\def\lapprox{\lower .7ex\hbox{$\;\stackrel{\textstyle <}{\sim}\;$}}
\def\gapprox{\lower .7ex\hbox{$\;\stackrel{\textstyle >}{\sim}\;$}}

\usepackage{color}
\definecolor{comment}{rgb}{0,0.3,0}
\definecolor{identifier}{rgb}{0.0,0,0.3}

\usepackage{listings}

\definecolor{listinggray}{gray}{0.9}
\definecolor{lbcolor}{rgb}{0.9,0.9,0.9}
\newcommand{\MSbar}{\overline{\mathrm{MS}}}

\begin{document}

\begin{figure}[h]
\epsfig{width=0.32\textwidth,figure=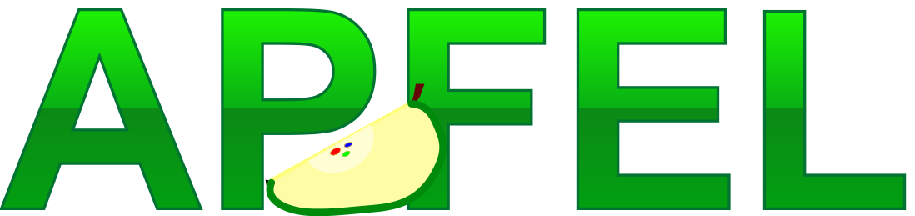}
\end{figure}

\vspace{-2.0cm}
\begin{flushright}
IFUM-1015-FT\\
CERN-PH-TH/2013-209\\
\end{flushright}
\vspace{1cm}

\begin{center}
{\Large \bf {\tt APFEL}: A PDF Evolution Library with QED corrections}
\vspace{.7cm}

Valerio~Bertone$^{1}$,
Stefano~Carrazza$^{2}$ and Juan~Rojo$^1$

\vspace{.3cm}
{\it ~$^1$ PH Department, TH Unit, CERN, CH-1211 Geneva 23, Switzerland \\
~$^2$ Dipartimento di Fisica, Universit\`a di Milano and
INFN, Sezione di Milano,\\ Via Celoria 16, I-20133 Milano, Italy\\
}
\end{center}

\vspace{0.1cm}

\begin{center}
{\bf \large Abstract}
\end{center}
Quantum electrodynamics and electroweak corrections are important ingredients for  many theoretical predictions 
at the LHC.
This paper documents {\tt APFEL}, a new PDF evolution
package that allows for the first time
 to perform DGLAP evolution up to NNLO in QCD and to LO
in QED, in the variable-flavor-number scheme and with either pole
or $\MSbar$ heavy quark masses.
 {\tt APFEL} consistently accounts for
the QED corrections to the evolution of quark and gluon
PDFs and
for the contribution from the photon PDF in the proton.
The  coupled  QCD$\otimes$QED 
  equations are solved in $x$-space by means
of higher order interpolation, followed by Runge-Kutta solution
of the resulting discretized evolution equations.
{\tt APFEL} is based on an innovative and flexible
 methodology for the sequential solution
of the QCD and QED evolution equations and their combination.
In addition to PDF evolution,  {\tt APFEL} provides a module that computes
Deep-Inelastic Scattering structure functions in the FONLL
general-mass variable-flavor-number 
scheme up to $\mathcal{O}\lp \alpha_s^2\rp$.
All the functionalities of {\tt APFEL} can be accessed via
a Graphical User Interface, supplemented with
a variety of  plotting tools for PDFs, parton
luminosities and structure functions.
Written in {\scshape Fortran 77}, {\tt APFEL} can also be used via the {\tt C/C++} and {\tt Python}
interfaces, and is publicly available from the {\tt HepForge} repository.

\newpage

\noindent {\Large \textbf{Program Summary}}\\

\noindent {\em Name of the program\/}: {\tt APFEL} \\[2mm]
{\em Version\/}: 2.0.0 \\[2mm]
{\em Program obtainable from\/}:
\url{http://projects.hepforge.org/apfel/}
\\[2mm]
{\em Distribution format\/}: compressed tar file and directly from the {\tt HepForge} svn repository \\[2mm]
{\em E-mail\/}: {\tt valerio.bertone@cern.ch}, {\tt stefano.carrazza@mi.infn.it} \,and {\tt juan.rojo@cern.ch} \\[2mm]
{\em License\/}: GNU Public License \\[2mm]
{\em Computers\/}: all \\[2mm]
{\em Operating systems\/}: all \\[2mm]
{\em Program language\/}: {\scshape Fortran~77}, {\tt C/C++} and {\tt Python} \\[2mm]
{\em Memory required to execute\/}:  $\lesssim$ 2 MB \\[2mm]
{\em Other programs called\/}: {\tt LHAPDF} \\[2mm]
{\em External files needed\/}: none \\[2mm]
{\em Number of bytes in distributed program, including test data
  etc.\/}: $\sim 2.4$~MB\\[2mm]
{\em Keywords\/}: unpolarised parton distribution functions (PDFs),
DGLAP evolution equations, QED  corrections, electroweak effects,
deep-inelastic scattering (DIS).
\\[2mm]
{\em Nature of the physical problem\/}: Solution of the unpolarized 
coupled DGLAP
evolution equations up to NNLO in QCD and to LO in QED in the variable-flavor-number scheme, both with pole and  with
$\MSbar$ masses.
\\[2mm]
{\em Method of solution\/}: Representation of parton distributions and
splitting functions on a grid in $x$, discretization of DGLAP
evolution equations and higher-order interpolation for general values
of $x$, numerical solution of the resulting discretized evolution
equations using Runge-Kutta methods.
\\[2mm]
{\em Restrictions on complexity of the problem\/}: Smoothness
of the initial conditions for the PDF evolution.
\\[2mm]
{\em Typical running time\/}: a few seconds for initialisation, then
$\sim$0.5~s for the generation of the PDF tables with combined QCD$\otimes$QED 
evolution (on a
Intel(R) Core(TM)2 Duo CPU E6750 @ 2.66GHz).

\clearpage

\tableofcontents

\clearpage

\section{Introduction}
\label{sec-intro}


The requirements of precision physics at the LHC demand parton
distribution functions (PDFs) that are determined using 
NLO and NNLO QCD theory (see~\cite{Forte:2013wc,Ball:2012wy,DeRoeck:2011na} for recent reviews).
For a substantial number of processes, the accuracy 
in both theoretical predictions and experimental
data is such that Quantum Electrodynamics (QED) and pure
 electroweak corrections also need to be included. Predictions for hadron-collider processes that include  QED and  electroweak corrections 
are available for
inclusive $W$ and $Z$ production~\cite{Baur:1998kt,Zykunov:2001mn,Dittmaier:2001ay,Baur:2001ze,Baur:2004ig,Arbuzov:2007db,Arbuzov:2005dd,Brensing:2007qm,Balossini:2009sa,CarloniCalame:2007cd,Dittmaier:2009cr}, $W$ and
$Z$ boson production in association with jets~\cite{Denner:2009gj,Denner:2011vu,Denner:2012ts}, diboson production~\cite{Baglio:2013toa,Bierweiler:2012kw,Luszczak:2013ata},
dijet production~\cite{Moretti:2005ut,Dittmaier:2012kx} and top quark pair production~\cite{Bernreuther:2005is,Kuhn:2005it,Hollik:2007sw,Hollik:2011ps,Kuhn:2013zoa} among others, see also Ref.~\cite{Mishra:2013una} for a recent review.

In order to derive consistent predictions for hadronic cross-sections,
both hard-scattering matrix elements and PDFs need to be determined
with the same accuracy in the QCD and electroweak couplings.
In particular, combining QCD and electroweak calculations at hadron colliders requires
parton distributions that have been determined using the coupled QCD$\otimes$QED 
DGLAP evolution equations~\cite{DeRujula:1979jj,Kripfganz:1988bd,Blumlein:1989gk}.
Until recently, the only PDF set which included QED corrections
was the  MRST04QED set, presented almost ten years ago~\cite{Martin:2004dh}.
Recently, the NNPDF2.3QED set~\cite{Ball:2013hta,Carrazza:2013bra,Carrazza:2013wua}
has also become available:
on top of providing an up-to-date PDF set determined using LO/NLO/NNLO QCD supplemented by LO QED theory, for the first time the photon PDF, 
with the corresponding 
uncertainty, has been extracted from LHC measurements on electroweak gauge
boson production, rather than being based on model assumptions.\footnote{The NNPDF2.3
QCD$\otimes$QED sets are available from {\tt LHAPDF}~\cite{Bourilkov:2006cj} 
starting from v5.9.0, as well
as internal PDF sets in {\tt Pythia8}~\cite{Sjostrand:2007gs}, see Ref.~\cite{Carrazza:2013axa} for details about the latter implementation.}

While much work has been devoted to the study of numerical solutions
of the QCD DGLAP evolution equations and their implementation in public
tools~\cite{Salam:2008qg,Cafarella:2008du,Botje:2010ay,Ratcliffe:2000kp,Schoeffel:1998tz,Pascaud:2001bi,pegasus,Kosower:1997hg}, 
less effort has been devoted to the solutions of the
DGLAP equations in the presence of
 QED corrections~\cite{Spiesberger:1994dm,Roth:2004ti,Martin:2004dh}.
To the best of our knowledge, the only public code which offers
such possibility  is {\tt partonevolution}~\cite{Weinzierl:2002mv,Roth:2004ti}.
However, {\tt partonevolution} is limited to NLO QCD corrections and in
 addition it does not allow
to explore different possibilities for the combination of the 
QCD and QED evolution equations.

Therefore, the main motivation for this work is to provide for
the first time
a public code, accurate and flexible, that can be used to
perform PDF evolution up to 
NNLO in QCD and LO in QED, both in the fixed-flavour-number (FFN) and
in the variable-flavour-number (VFN) schemes, and using either pole or $\MSbar$ heavy quark
masses. 
Another motivation was to complete
the discussion about the solutions of the
QCD$\otimes$QED combined equations that was only sketched in the 
original 
NNPDF2.3QED publication~\cite{Ball:2013hta}, as well as to provide
another independent validation of the {\tt FastKernel} implementation
used in the determination of those PDF sets.

We call this code {\tt APFEL}, which stands
for {\it A Parton distribution Function Evolution Library}.
{\tt APFEL} is based on an innovative methodology for the solution
of the QCD and QED evolution equations and their combination, leading
to a flexibility that can be used to explore
various options that differ by subleading terms.
{\tt APFEL} solves the QCD$\otimes$QED equations sequentially, that is,
 first
performing QCD evolution and then QED evolution, or
vice-versa. 
The particular ordering, as well as the related choices to be made when crossing
heavy quark thresholds, can be easily modified by the user.
These various possibilities differ only by subleading terms, that can
however lead to a phenomenological impact, so it is important
to quantify  these in some detail.

Public codes for the evolution of parton distributions and 
fragmentation functions can be divided according to the method
they use to solve the DGLAP equations.
A first family consists of $x$-space methods, which typically use
a representation of the PDFs on a grid in $x$ together
with higher-order interpolation techniques for the solution
of the intergro-differential equations~\cite{Salam:2008qg,Botje:2010ay,Schoeffel:1998tz,Pascaud:2001bi,Ratcliffe:2000kp}. 
The widely used {\tt HOPPET}~\cite{Salam:2008qg} and {\tt QCDNUM}~\cite{Botje:2010ay} 
programs belong to this family.
The other family is composed of $N$-space codes, where the
DGLAP equations are  first transformed into Mellin space,
analytically solved, and then inverted back to $x$-space
using complex-variable methods~\cite{pegasus,Cafarella:2008du,Weinzierl:2002mv,Roth:2004ti,Kosower:1997hg}. 
The
{\tt PEGASUS}~\cite{pegasus} program is one of the best-known examples
of this strategy.
 The main drawback of the $N$-space methods, however, is the fact that they
require the analytical Mellin transform of the initial PDFs which is
possible only for some very specific functional forms.
A third approach is provided by the hybrid method adopted in the 
{\tt FastKernel} program,  the internal code used in the NNPDF fits~\cite{DelDebbio:2007ee,Ball:2008by}, where
the DGLAP equations are solved in Mellin space and then used to determine
the $x$-space evolution operators, which are convoluted
with the $x$-space PDFs to perform the evolution.

Inspired in part by the techniques used in the  {\tt HOPPET} 
and {\tt QCDNUM} programs, {\tt APFEL} solves the 
 QCD$\otimes$QED  evolution
 equations in $x$-space by means
of higher-order interpolation, followed by Runge-Kutta solution
of the resulting discretized evolution equations.
Its main strength is that, on top of providing fast and efficient state-of-the-art
QCD evolution, it allows to use the variable-flavour-number
 scheme in the QED evolution of PDFs and 
to systematically explore a range of
possible options to solve the combined evolution equations.
For these reasons, we believe that {\tt APFEL}
has the potential to become an
 important tool for the present and coming generation of
 PDF analyses including QED corrections.

On top of the PDF evolution routines, {\tt APFEL} includes also
two additional modules that should be of interest for a variety of
users.
The first one allows the computation of neutral- and charged-current
Deep-Inelastic Scattering (DIS) structure functions up to
 $\mathcal{O}\lp \alpha_s^2\rp$ in the FONLL general-mass
VFN scheme~\cite{Forte:2010ta} as well as in the 
zero-mass VFN and FFN schemes~\cite{Ball:2013gsa}.
The second is a user-friendly, flexible Graphical User Interface
(GUI),  which provides easy access to all the functionalities of
{\tt APFEL}, and that is supplemented with a variety of 
graphical plotting tools
for PDFs, parton luminosities and structure functions.

The outline of this paper is the following.
 In Sect.~\ref{sec-theory}
we review the structure of the coupled QCD$\otimes$QED
evolution equations and discuss their solution
and  the different options for the treatment
of subleading terms, and we also briefly discuss the implementation 
deep-inelastic structure functions.
In Sect.~\ref{sec:methods} we describe the numerical
techniques that are used in {\tt APFEL}.
Then
in Sect.~\ref{sec-manual} we introduce the main
functionalities of {\tt APFEL} and describe the 
standard user interface,
as well as the Graphical User
Interface which provides a user-friendly interface to
all these functionalities and to a variety of
plotting tools.
In Sect.~\ref{sec-benchmarks} we validate {\tt APFEL} by benchmarking it
against other publicly available tools, and finally in Sect.~\ref{sec-conclusion}
we conclude and discuss possible future developments.

\section{DGLAP evolution with QED corrections}
\label{sec-theory}

In this section we present the strategy that {\tt APFEL}
adopts in order to perform the
DGLAP evolution of parton distributions when
both QCD and QED effects are taken into account.
First of all, we will present the method used to solve
the QED evolution equations, then how to combine the
QCD and QED solutions, and finally the treatment of
 heavy quark thresholds in the variable-flavor-number scheme.
QCD corrections to the DGLAP evolution equations~\cite{ap,gl,dok} 
are available up to NNLO~\cite{gNLOa,gNLOb,gNLOc,gNLOe,gNLOf,gNNLOa,gNNLOb,Buza:1995ie}, and the structure of their
solutions has been discussed in great detail in the
literature, see for instance~\cite{Salam:2008qg,pegasus,Forte:2013wc} and references therein.
In this section we limit ourselves to discussing only
the new features that arise in the DGLAP equations
 when QED effects are taken into account.
Finally, in the last subsection we briefly review the theory of
Deep-Inelastic Scattering structure functions.

\subsection{Solving the QED evolution equations}
\label{SolvingQEDEquation}

The implementation of the QED corrections to the
DGLAP evolution equations leads to the inclusion of additional terms which
contain QED splitting 
functions~\cite{DeRujula:1979jj,Kripfganz:1988bd,Blumlein:1989gk}, proportional to the QED coupling
$\alpha$, convoluted with the PDFs.
There are several possibilities to solve the coupled QCD$\otimes$QED DGLAP
evolution equations, and, as opposed to previous works,
  {\tt APFEL} adopts a fully factorized approach. 
In this approach, the QCD and the QED factorization procedures can be
regarded as two independent steps that lead to two independent factorization scales,
on which all PDFs depend, denoted by $\mu$ for QCD and $\nu$ for QED,
that is, $q_i\equiv q_i(x,\mu,\nu)$.

In the particular case where  QED
corrections are included up to
$\mathcal{O}(\alpha)$ and the mixed subleading terms
$\mathcal{O}(\alpha\alpha_s)$ are neglected, the QCD evolution with
respect to $\mu$ and the QED evolution with respect to $\nu$ will
be given by two fully decoupled equations:
\begin{equation}\label{DGLAPequationsQCD_QED}
\begin{array}{rcl}
\displaystyle \mu^{2}\frac{\partial}{\partial \mu^{2}}{\mathbf q}(x,\mu,\nu) &=&
\displaystyle {\mathbf P}^{\rm QCD} (x,\alpha_s(\mu))\otimes {\mathbf
q}(x,\mu,\nu)\,,\\
\\
\displaystyle \nu^{2}\frac{\partial}{\partial \nu^{2}}{\mathbf
 q}(x,\mu,\nu) &=&
\displaystyle {\mathbf P}^{\rm QED} (x,\alpha(\nu))\otimes {\mathbf
q}(x,\mu,\nu)\,,
\end{array}
\end{equation}
where ${\mathbf P}^{\rm QCD}$ and ${\mathbf P}^{\rm QED}$ are respectively 
the QCD and
QED matrices of splitting functions and 
${\mathbf q}(x,\mu,\nu)$ is a vector
containing all the parton distribution functions.
Let us recall that in the presence of QED corrections, 
the photon PDF $\gamma(x,\mu,\nu)$
should also be included in ${\mathbf q}(x,\mu,\nu)$.
The symbol $\otimes$ in Eq.~(\ref{DGLAPequationsQCD_QED}) represents
the usual convolution operator defined as:
\begin{equation}
A(x)\otimes B(x) \equiv \int_0^1 dy \int_0^1 dz\,A(y)B(z) \delta(x-yz)\,.
\end{equation}
The independent solutions of the differential equations in
Eq.~(\ref{DGLAPequationsQCD_QED}), irrespective of the numerical
technique used, will give as a result two different
evolution operators: ${\bm \Gamma}^{\rm QCD}$, that evolves the array
 $\mathbf q$ in $\mu$ while keeping $\nu$ constant, and ${\bm \Gamma}^{\rm QED}$,
that evolves $\mathbf q$ in $\nu$ while keeping $\mu$ constant. 
If the QCD evolution takes place between $\mu_0$ and $\mu_1$ and
the QED evolution
between $\nu_0$ and $\nu_1$, we will have that:
\begin{equation}\label{DecoupledDGLAPequations}
\begin{array}{l}
\displaystyle {\mathbf q}(x,\mu_1,\nu_{\rm
 })={\bm \Gamma}^{\rm QCD}(x|\mu_1,\mu_0)\otimes{\mathbf q}(x,\mu_0,\nu_{\rm })\,,\\
\\
\displaystyle {\mathbf q}(x,\mu_{\rm
   },\nu_1)={\bm \Gamma}^{\rm QED}(x|\nu_1,\nu_0)\otimes{\mathbf q}(x,\mu_{\rm },\nu_0)\,.\\
\end{array}
\end{equation}
Once the QCD and QED evolution operators in Eq.~(\ref{DecoupledDGLAPequations}) have been
calculated, one can combine them to obtain a coupled
evolution operator ${\bm \Gamma}^{\rm QCD \otimes QED}$ that evolves PDFs
both in the QCD and in the QED scales, that is:
\begin{equation}
{\mathbf q}(x,\mu_1,\nu_1)={\bm \Gamma}^{\rm QCD \otimes QED}(x|\mu_1,\mu_0;\nu_1,\nu_0)\otimes{\mathbf
q}(x,\mu_0,\nu_0)\,.
\end{equation}
The derivation of the combined evolution operator  ${\bm \Gamma}^{\rm QCD \otimes QED}$ will be discussed in Sect.~\ref{sec:CombinedQCDandQED}.

Let us present first the strategy used in {\tt APFEL} to solve the QED DGLAP equations
in Eq.~(\ref{DGLAPequationsQCD_QED}).\footnote{In what concerns the QCD evolution equations,
 {\tt APFEL} implements a similar strategy as the one used in~\cite{Ball:2008by}, namely  rotating the
PDF vector $\mathbf q$ from the flavor
basis into the evolution basis, the one which
maximally diagonalizes the QCD splitting
function matrix ${\mathbf P}^{\rm QCD}$.
Then the  QCD DGLAP
equations in this evolution basis are solved using the numerical techniques
 that will be presented in Sect.~\ref{sec:methods}.}
At leading order, the QED equations for the evolution of the quark and photon
PDFs, dropping for simplicity the dependence on the QCD factorization scale $\mu$, read:
\begin{equation}\label{QED_DGLAP}
\begin{array}{rcl}
\displaystyle \nu^{2}\frac{\partial}{\partial \nu^{2}}\gamma(x,\nu)
&=& \displaystyle \frac{\alpha(\nu)}{4\pi} \left[\left(\sum_{i}N_c e_{i}^{2}\right)
P_{\gamma\gamma}^{(0)}(x)\otimes\gamma(x,\nu)+\sum_ie_{i}^{2}P_{\gamma
  q}^{(0)}(x)\otimes (q_{i}+\bar{q}_{i})(x,\nu)\right]\,,\\
\\
\displaystyle \nu^{2}\frac{\partial}{\partial
   \nu^{2}} q_{i}(x,\nu)&=&\displaystyle \frac{\alpha(\nu)}{4\pi}
 \left[ N_c e_{i}^{2} P_{q\gamma}^{(0)}(x)\otimes \gamma(x,\nu)+e_{i}^{2}
 P_{qq}^{(0)}(x)\otimes q_i(x,\nu)\right]\,,\\
\\
\displaystyle \nu^{2}\frac{\partial}{\partial
   \nu^{2}} \bar{q}_{i}(x,\nu)&=&\displaystyle \frac{\alpha(\nu)}{4\pi}
 \left[ N_c e_{i}^{2} P_{q\gamma}^{(0)}(x)\otimes \gamma(x,\nu)+e_{i}^{2}
 P_{qq}^{(0)}(x)\otimes \bar{q}_i(x,\nu)\right]\,, 
\end{array}
\end{equation}
where $\gamma(x,\nu)$, $q_i(x,\nu)$ and $\bar{q}_i(x,\nu)$ are
respectively the PDFs of the photon, the $i$-th quark and
the $i$-th antiquark, $e_i$ the quark electric charge, $N_c=3$ the number
of colors and $\alpha(\nu)$ the running fine structure constant.
Note that at this order  the gluon PDF does not enter the
QED evolution equations.
The leading-order QED splitting functions $P_{ij}(x)$ are given by:
\begin{equation}\label{QEDLOsplittingFunctions}
\begin{array}{rcl}
\displaystyle P_{q\gamma}^{(0)}(x) &=& \displaystyle 2\left[x^2+(1-x)^2\right],\\
\\
\displaystyle P_{\gamma q}^{(0)}(x) &=& \displaystyle
2\left[\frac{1+(1-x)^2}{x}\right],\\
\\
\displaystyle P_{\gamma\gamma}^{(0)}(x) &=& \displaystyle
-\frac{4}{3}\delta(1-x),\\
\\
\displaystyle P_{qq}^{(0)}(x) &=& \displaystyle
2\frac{1+x^2}{(1-x)_+}+3\delta(1-x)\,.
\end{array}
\end{equation}
The index $i$ in Eq.~(\ref{QED_DGLAP}) runs over the active quark flavors
at a given scale $\nu$. 

It should be noted that, in the presence of QED effects, the usual
momentum sum rule is modified to take into account the contribution
coming from the photon PDF. Therefore, provided that
the input PDFs respect the momentum sum rule, the QED evolution should
satisfy the equality:
 \begin{equation}
 \int_{0}^{1}dx~x\left\{ \underset{i}{\sum}(q_{i}+\bar{q}_{i})(x,\mu,\nu)+g(x,\mu,\nu)+\gamma(x,\mu,\nu)\right\} =1 \ ,
 \label{eq:momsr}
 \end{equation}
for any value of the scales $\mu$ and $\nu$. An important
test of the numerical implementation of DGLAP evolution in the
presence of QED effects is to check that Eq.~(\ref{eq:momsr}) indeed
holds at all scales. 

As in the case of QCD, an important practical issue 
that needs to be addressed
when solving the QED DGLAP evolution
equations is the choice of the PDF basis. 
The use of the flavor basis ${\mathbf q} =\{\gamma,u,\bar{u},d,\bar{d},...\}$
requires the solution of a system of thirteen coupled equations which in turns 
leads to a cumbersome numerical implementation.  
This problem can be overcome by choosing a suitable PDF
basis, the evolution basis, that maximally diagonalizes the
QED splitting function matrix.
Note that this optimized basis  will be different from that used in QCD, due to
the presence of the electric charges $e_i$ in Eq.~(\ref{QED_DGLAP}) that are different between up- and
down-type quarks.\footnote{ This difference between up- and
  down-type quarks, in the presence of QED effects, is also responsible for the dynamical generation of isospin symmetry breaking between
proton and neutron PDFs.}

In {\tt APFEL} we adopt a PDF basis for the QED evolution which was originally suggested in
Ref.~\cite{Roth:2004ti}, defined by the following
singlet and non-singlet PDF combinations:
\begin{equation}\label{QEDEvolutionBasis}
\begin{array}{rcl}
\text{Singlet}: &\quad& \mathbf{q}^{\rm SG}=
\begin{pmatrix}
\gamma\\
\Sigma\equiv u^{+}+c^{+}+t^{+}+d^{+}+s^{+}+b^{+}\\
D_{\Delta\Sigma} \equiv u^{+}+c^{+}+t^{+}-d^{+}-s^{+}-b^{+}
\end{pmatrix}\,,\\
\\
\text{ Non-Singlet}: &\quad& q_i^{\rm NS}=\left\{\begin{array}{c}
D_{uc}\equiv u^{+}-c^{+},\\
D_{ds} \equiv  d^{+}-s^{+},\\
D_{sb} \equiv s^{+}-b^{+},\\
D_{ct} \equiv c^{+}-t^{+},\\
u^{-},\\
d^{-},\\
s^{-},\\
c^{-},\\
b^{-},\\
t^{-}
\end{array}\right\}\,,\quad i = 1,\dots,10\,,
\end{array}
\end{equation}
where we have defined $q^{\pm} \equiv q \pm \overline{q}$.
Similarly to the QCD notation, the singlet distributions are those
that couple to the  photon PDF $\gamma(x,\nu)$, while 
the non-singlet distributions evolve multiplicatively and do not
couple to the photon.

With the choice of basis of  Eq.~(\ref{QEDEvolutionBasis}), the
original thirteen-by-thirteen system of
 coupled equations in the flavor basis reduce to a 
three-by-three system of coupled equations and ten additional 
decoupled differential equations.
Expressing the QED DGLAP equations given in Eq.~(\ref{QED_DGLAP}) in
terms of this evolution basis, 
we find that the singlet PDFs evolve as follows:
\begin{equation}\label{SingletEvolution}
\nu^{2}\frac{\partial}{\partial \nu^{2}}
\begin{pmatrix}
\gamma\\
\Sigma\\
D_{\Delta\Sigma}
\end{pmatrix}=\frac{\alpha(\nu)}{4\pi}
\begin{pmatrix}
e_{\Sigma}^{2}P_{\gamma\gamma}^{(0)} & \eta^{+}P_{\gamma q}^{(0)} & \eta^{-}P_{\gamma q}^{(0)}\\
\theta^{-}P_{q\gamma}^{(0)} & \eta^{+}P_{qq}^{(0)} & \eta^{-}P_{qq}^{(0)}\\
\theta^{+}P_{q\gamma}^{(0)} & \eta^{-}P_{qq}^{(0)} & \eta^{+}P_{qq}^{(0)}
\end{pmatrix}\otimes
\begin{pmatrix}
\gamma\\
\Sigma\\
D_{\Delta\Sigma}
\end{pmatrix}\,,
\end{equation}
where, using the fact that $e_{u}^{2}=e_{c}^{2}=e_{t}^{2}$ and
$e_{d}^{2}=e_{s}^{2}=e_{b}^{2}$, we have defined:
\begin{equation}
\begin{array}{rcl}
e_{\Sigma}^{2}& \equiv &\displaystyle
N_c(n_{f,\text{up}}e_{u}^{2}+n_{f,\text{dn}}e_{d}^{2})\,,\\
\\
\eta^{\pm} & \equiv & \displaystyle \frac{1}{2}\left(e_{u}^{2}\pm
  e_{d}^{2}\right)\,,\\
\\
\theta^{\pm} & \equiv & \displaystyle
2N_{c}n_{f}\left[\left(\frac{n_{f,\text{up}}-n_{f,\text{dn}}}{n_{f}}\right)\eta^{\pm}+\eta^{\mp}\right]\,,
\end{array}
\end{equation}
where $n_{f,\text{up}}$ and $n_{f,\text{dn}}$ are the number of up-
and down-type active quark flavors, respectively, and $n_{f}=n_{f,\text{up}}+n_{f,\text{dn}}$.
The non-singlet PDFs, instead, obey the multiplicative evolution equation:
\begin{equation}\label{NonSingletEvolution}
\nu^{2}\frac{\partial}{\partial \nu^{2}} q_i^{\rm NS}(x,\nu) = e_i^2 P_{qq}^{(0)}(x)
\otimes q_i^{\rm NS}(x,\nu)\,,
\end{equation}
where the electric charge $e_i^2 = e_u^2$ for the up-type
distributions $q_i^{\rm NS}  = D_{uc},
D_{ct}, u^-,c^-,t^-$ while $e_i^2 = e_d^2$ for the down-type
distributions $q_i^{\rm NS}  = D_{ds},
D_{sb}, d^-,s^-,b^-$. 
Let us mention  that strictly speaking
  Eq.~(\ref{NonSingletEvolution}) is valid only if all the quark
  flavors are present in the evolution, that is for $n_f=6$. 
For $3 \le n_f \le 5$, some non-singlet PDF ($D_{uc}$, $D_{sb}$ and
$D_{ct}$) will not evolve
independently, since they can be written
as a linear combination of
singlet PDFs.
For instance, below the charm threshold, $D_{uc}=u^+=(\Sigma+D_{\Delta\Sigma})/2$.

The solution of Eqs.~(\ref{SingletEvolution}) and~(\ref{NonSingletEvolution}) determines
 the QED evolution operators that
evolve the singlet and non-singlet PDFs from the initial scale $\nu_0$
to some final scale $\nu$ according to the equations:
\begin{equation}\label{SolutionQED}
\begin{array}{rcl}
{\mathbf q}^{\rm SG}(x,\nu) &=& {\bm \Gamma}^{\rm SG}_{\rm QED}(x|\nu,\nu_0)
\otimes {\mathbf q}^{\rm SG}(x,\nu_0)\,,\\
\\
q_i^{\rm NS}(x,\nu) &=& \Gamma_{{\rm QED},i}^{\rm NS}(x|\nu,\nu_0) \otimes q_i^{\rm NS}(x,\nu_0)\,,
\end{array}
\end{equation}
where the singlet evolution operator ${\bm \Gamma}^{\rm SG}_{\rm QED}$ is a three-by-three
matrix while the non-singlet evolution operators $\Gamma_{{\rm QED},i}^{\rm NS}$
form an scalar array.
In Sect.~\ref{sec:methods}
we will show how to compute numerically
these evolution operators solving the corresponding integro-differential equations by means of
higher-order interpolation techniques.

\subsection{Combining the QCD and QED evolution operators}
\label{sec:CombinedQCDandQED}

Once the QED evolution operators in Eq.~(\ref{SolutionQED}) have
been computed by means of some suitable numerical method,
one needs to combine them with the corresponding QCD
evolution operators.
In order to perform the combination, we
can write Eq.~(\ref{SolutionQED}) in a matrix form introducing in the
PDF basis also the gluon PDF  $g(x,\nu,\mu)$. 
Taking into
account the fact that at leading order in QED the gluon PDF does not evolve,
reintroducing the dependence on the QCD factorization scales $\mu$ and
dropping for simplicity the dependence on $x$, we 
can write  Eq.~(\ref{SolutionQED}) as follows:
\begin{equation}\label{SolutionQEDMatr}
\underbrace{\begin{pmatrix}
g(\mu,\nu)\\
{\mathbf q}^{\rm SG}(\mu,\nu)\\
q_1^{\rm NS}(\mu,\nu)\\
\vdots \\
q_{10}^{\rm NS}(\mu,\nu)\\
\end{pmatrix}}_{{\mathbf q}^{}(\mu,\nu)} =
\underbrace{\begin{pmatrix}
1 & 0 & 0 & 0 & 0\\
0 & {\bm \Gamma}^{\rm SG}_{\rm QED} & {0} & \dots & {0} \\
0 & {0}  & \Gamma_{{\rm QED},1}^{\rm NS} & \dots & 0 \\
\vdots &  \vdots  & \vdots & \ddots & \vdots \\
0 & {0}  & 0 & \dots   & \Gamma_{{\rm QED},10}^{\rm NS}
\end{pmatrix}}_{{\bm \Gamma}^{\rm QED}(\nu,\nu_0)}
\otimes
\underbrace{\begin{pmatrix}
g(\mu,\nu_0)\\
{\mathbf q}^{\rm SG}(\mu,\nu_0)\\
q_1^{\rm NS}(\mu,\nu_0)\\
\vdots \\
q_{10}^{\rm NS}(\mu,\nu_0)\\
\end{pmatrix}}_{{\mathbf q}^{\rm }(\mu,\nu_0)}\,.
\end{equation}
In the above expression, we have denoted by ${\mathbf q}^{\rm }(\mu,\nu)$ the fourteen-dimensional
vector that contains all PDF combinations in the QED evolution basis 
of Eq.~(\ref{QEDEvolutionBasis}) plus the gluon PDF.
Of course, a similar expression as that of Eq.~(\ref{SolutionQED})
will hold for the solution of the QCD DGLAP evolution equations:
\begin{equation}\label{SolutionQCDMatr1}
\widetilde{\mathbf q}^{\rm }(\mu,\nu) = \widetilde{\bm \Gamma}^{\rm QCD}(\mu,\mu_0) \otimes \widetilde{\mathbf q}^{\rm }(\mu_0,\nu)\,,
\end{equation}
where in this case the vector ${ \widetilde{\mathbf q} }$ is given in the
QCD evolution basis, which is a different
linear combination of the quark, anti-quark, gluon and photon
PDFs as compared to the
corresponding QED evolution basis.
The two basis are related by an invertible fourteen-by-fourteen 
rotation matrix
$\mathbf T$ that transforms the vector $\widetilde{\mathbf q}^{\rm }$
into the vector ${\mathbf q}^{\rm }$:
\begin{equation}\label{TransformationQCDxQED}
{\mathbf q}^{\rm } = {\mathbf T}\cdot \widetilde{\mathbf q}^{\rm } \quad
\Longrightarrow \quad \widetilde{\mathbf q}^{\rm } = {\mathbf T}^{-1}\cdot {\mathbf q}^{\rm }\,.
\end{equation}
Using Eq.~(\ref{TransformationQCDxQED}) and the condition ${\mathbf
  T}\cdot{\mathbf T}^{-1} = {\bm 1}$, the solution of
the QED evolution equations Eq.~(\ref{SolutionQEDMatr}) can
be rotated as follows:
\begin{equation}\label{SolutionQEDMatr1}
\widetilde{\mathbf q}^{\rm }(\mu,\nu) = \underbrace{\left[{\mathbf
  T}^{-1}\cdot {\bm \Gamma}^{\rm QED}(\nu,\nu_0) \cdot {\mathbf T}\right]}_{\widetilde{{\bm \Gamma}}^{\rm QED}(\nu,\nu_0)}
\otimes ~\widetilde{\mathbf q}^{\rm }(\mu,\nu_0)\,.
\end{equation}
where $\widetilde{{\bm \Gamma}}^{\rm QED}(\nu,\nu_0)$ is now the QED evolution operator
expressed in the QCD evolution basis.
Eqs.~(\ref{SolutionQCDMatr1}) and~(\ref{SolutionQEDMatr1}) determine
 the QCD and the QED evolution, respectively, of PDFs in the QCD evolution basis
 and can therefore be consistently used to construct a
combined   QCD$\otimes$QED
evolution operator.
In the following, we drop all the tildes since it is understood
that PDFs and evolution operators are always expressed in the
QCD evolution basis.

Now, when combining QCD and QED evolution operators 
we are faced with an inherent ambiguity.
Given that QCD and QED evolutions
take place by means of the matrix evolution operators
${\bm \Gamma}^{\rm QCD}$ and ${{\bm \Gamma}}^{\rm QED}$ that
 do not commute,
\begin{equation}\label{commutator}
[{\bm \Gamma}^{\rm QCD},{{\bm \Gamma}}^{\rm QED}] \neq 0\,,
\end{equation}
this implies  that performing first the QCD evolution followed by the
QED evolution leads to a different result if the opposite order is
assumed. 
We can then define the two possible cases:
\begin{equation}
\label{eq:QCED}
{\bm \Gamma}^{\rm QCED}(\mu,\mu_0;\nu,\nu_0) \equiv {{\bm
    \Gamma}}^{\rm QED}(\nu,\nu_0)\otimes{\bm \Gamma}^{\rm
  QCD}(\mu,\mu_0)\,,
\end{equation}
\begin{equation}
\label{eq:QECD}
{\bm \Gamma}^{\rm QECD}(\mu,\mu_0;\nu,\nu_0) \equiv {\bm \Gamma}^{\rm QCD}(\mu,\mu_0)\otimes {{\bm \Gamma}}^{\rm QED}(\nu,\nu_0)\,,
\end{equation}
and the condition in Eq.~(\ref{commutator}) implies that:
\begin{equation}
\label{eq:diff}
{\bm \Gamma}^{\rm QCED}(\mu,\mu_0;\nu,\nu_0) \otimes {\mathbf q}^{\rm
}(\mu_0,\nu_0) \neq {\bm \Gamma}^{\rm QECD}(\mu,\mu_0;\nu,\nu_0) \otimes {\mathbf q}^{\rm
}(\mu_0,\nu_0)\,.
\end{equation}
However, using the analytical solution of the QCD and QED DGLAP equations in Mellin space and
the Baker-Campbell-Hausdorff formula, it is possible to show that:
\begin{equation}\label{commutator1}
[{\bm \Gamma}^{\rm QCD},{{\bm \Gamma}}^{\rm QED}] = \mathcal{O}(\alpha\alpha_s)\,,
\end{equation}
and therefore Eqs.~(\ref{eq:QCED}) and~(\ref{eq:QECD}) correspond
to the same evolution operator up to perturbative subleading $\mathcal{O}(\alpha\alpha_s)$ terms,
which are beyond the accuracy of the present implementation of QED effects.

In addition, a  careful analysis of the expansions of the two combined
evolution operators in Eqs.~(\ref{eq:QCED}) and~(\ref{eq:QECD}) shows
that they have a similar perturbative structure:
\begin{equation}\label{ExpansionQCED}
{\bm \Gamma}^{\rm QCED} = \sum_{n=0}^{\infty} \left(\alpha {\bm A} + \alpha_s {\bm B}\right)^n +
\alpha\alpha_s{\bm C} + \mathcal{O}(\alpha^2)\,,
\end{equation}
\begin{equation}\label{ExpansionQECD}
{\bm \Gamma}^{\rm QECD} = \sum_{n=0}^{\infty} \left(\alpha {\bm A} + \alpha_s {\bm B}\right)^n -
\alpha\alpha_s{\bm C} + \mathcal{O}(\alpha^2)\,.
\end{equation}
These expansions suggest a third possibility for the  combined
evolution operator given by the average of the ${\bm \Gamma}^{\rm QCED}$
and ${\bm \Gamma}^{\rm QECD}$ operators:
\begin{equation}\label{AveragedSolution}
{\bm \Gamma}^{\rm QavD} \equiv \frac{{\bm \Gamma}^{\rm QCED} + {\bm \Gamma}^{\rm QECD}}{2}\,,
\end{equation}
so that the subleading terms $\mathcal{O}(\alpha\alpha_s)$ cancel and
the perturbative remainder is  $\mathcal{O}(\alpha^2)$.

As will be shown in Sect.~\ref{sec-benchmarks}, the {\tt QavD} solution,
Eq.~(\ref{AveragedSolution}), turns out to be 
the closest to the solution of the QCD$\otimes$QED equations adopted
in the MRST04QED fit~\cite{Martin:2004dh} and
in {\tt partonevolution}~\cite{Weinzierl:2002mv,Roth:2004ti}, 
all of them different by  $\mathcal{O}(\alpha^2)$ terms only.
In Sect.~\ref{sec-benchmarks} we will also study
 the  numerical impact of the different options for 
computing the QCD$\otimes$QED
evolution operators, Eqs.~(\ref{eq:QCED},\ref{eq:QECD},\ref{AveragedSolution}).
There we will show that the subleading terms
in solutions {\tt QCED} and {\tt QECD} are numerically sizable because they are
enhanced by large unresummed scale logarithms, but that
this is not the case for the {\tt QavD} solution.

Let us mention also that, to the best of our knowledge, this is the first
time that the sequential combination of the QCD and QED
evolution has been investigated in the literature.
Programs such as {\tt partonevolution} instead diagonalize the sum of the
QCD and the QED splitting matrices in a special basis, rather than solving the QCD and the QED 
DGLAP evolution equations separately and then combining the results, as {\tt APFEL} does.

\subsection{QCD$\otimes$QED combined evolution in the VFN scheme}
\label{sec:vfns}

The above discussion assumed that no heavy quark threshold is crossed
during the DGLAP evolution, that is, it is valid only when
PDF evolution is performed 
in the FFN scheme.
Of course, for any realistic application we need to perform
PDF evolution in the VFN scheme, 
where the number of active quark flavors $n_f$ increases by
one each time a heavy quark mass threshold is crossed.
From the practical point of view, solving the evolution
equations in the VFN scheme implies solving different evolution
equations below and above each heavy quark threshold and matching them
at the threshold itself, as we discuss now.

In order to show how {\tt APFEL} performs the QCD$\otimes$QED combined
evolution in the VFN scheme, we use as an example the crossing of the
charm mass threshold $m_c$ (\textit{i.e.} $\mu_0,\nu_0<m_c<\mu,\nu$)
where the number of active flavors contributing to the evolution
increases from three to four. 
In this case, the QCD and the QED evolution of
the PDF vector ${\mathbf q}^{\rm }$ can be schematically expressed
as follows:
\begin{equation}
\begin{array}{rcl}
{\mathbf q}^{\rm
}(\mu,\nu_0)&=&{\bm \Gamma}^{{\rm QCD},(4)}(\mu,m_c)\otimes {\bm \Gamma}^{{\rm QCD},(3)}(m_c,\mu_0)\otimes {\mathbf q}^{\rm
}(\mu_0,\nu_0)\,,\\
\\
{\mathbf q}^{\rm
}(\mu_0,\nu)&=&{{\bm \Gamma}}^{{\rm QED},(4)}(\nu,m_c)\otimes{{\bm \Gamma}}^{{\rm QED},(3)}(m_c,\nu_0)\otimes {\mathbf q}^{\rm
}(\mu_0,\nu_0)\,,
\end{array}
\end{equation}
where the upper index in the evolution operators ${\bm \Gamma}$ denotes
the number of active flavors. 
Now, there are different possibilities. Choosing for instance to
perform first the QCD followed by the QED evolution, we have two options:
\begin{equation}\label{ParallelAndSeries}
\begin{array}{rcl}
{\mathbf q}(\mu,\nu)&=&\bigg\{\left[{\bm \Gamma}^{{\rm
 QED},(4)}(\nu,m_c)\otimes{\bm \Gamma}^{{\rm
 QCD},(4)}(\mu,m_c)\right]\otimes\\
\\
&& \left[{\bm \Gamma}^{{\rm
 QED},(3)}(m_c,\nu_0)\otimes {\bm \Gamma}^{{\rm QCD},(3)}(m_c,\mu_0)\right]\bigg\}\otimes {\mathbf q}(\mu_0,\nu_0)\\
\\
&\equiv& {\bm \Gamma}^{\rm QCEDP}(\mu,\mu_0;\nu,\nu_0) \otimes {\mathbf q}(\mu_0,\nu_0)\,,\\
\\
{\mathbf q}(\mu,\nu)&=&\bigg\{\left[{\bm \Gamma}^{{\rm
 QED},(4)}(\nu,m_c)\otimes {\bm \Gamma}^{{\rm
 QED},(3)}(m_c,\nu_0)\right]\otimes\\
\\
&& \left[{\bm \Gamma}^{{\rm
 QCD},(4)}(\mu,m_c)\otimes {\bm \Gamma}^{{\rm QCD},(3)}(m_c,\mu_0)\right]\bigg\}\otimes {\mathbf q}(\mu_0,\nu_0)\\
\\
&\equiv& {\bm \Gamma}^{\rm QCEDS}(\mu,\mu_0;\nu,\nu_0) \otimes {\mathbf q}(\mu_0,\nu_0)\,.\\
\end{array}
\end{equation}
In the first of these equations, QCD and QED
evolutions are done in parallel (thus the notation ${\bm \Gamma}^{\rm  QCEDP}$),
that is, QCD and QED evolutions with three active flavors are performed
before crossing the charm threshold and then again
with four active flavors after the crossing. 
In the second equation in
Eq.~(\ref{ParallelAndSeries}), instead, QCD and QED evolutions are
done in series (${\bm \Gamma}^{\rm QCEDS}$), that is, the
full QCD evolution, including the crossing of
the charm threshold, is followed by the full QED one.

The same discussion applies when QED evolution is followed by QCD evolution,
defining the evolution operators ${\bm \Gamma}^{\rm QECDP}$ and ${\bm
  \Gamma}^{\rm QECDS}$, and to the averaged solution,
defining the evolution operators ${\bm \Gamma}^{\rm QavDP}$ and ${\bm
  \Gamma}^{\rm QavDS}$.
Again, due to  Eq.~(\ref{commutator1}), all the six possibilities are
formally equivalent up to subleading terms. 
They have all been implemented in {\tt APFEL} and in
Sect.~\ref{sec-benchmarks} we will study their numerical
differences. 
Let us finally 
mention  that since in practice there is no need to keep the QCD and the QED
factorization scales different, in 
{\tt APFEL} they are always taken to be equal, \textit{i.e.}
$\nu_0=\mu_0=Q_0$ and $\nu = \mu = Q$.

\subsection{Deep-Inelastic Scattering structure functions}
\label{sec-dis}

In this section we briefly review the theory of the Deep-Inelastic 
Scattering (DIS) structure functions, and describe
the options that are available in {\tt APFEL}. 
DIS cross-sections for neutral- and charged-current DIS on
unpolarized nucleons are given by the contribution of 
three independent structure functions,
that are usually taken to be $F_2$, $F_L$ and $xF_3$, 
so that one can generically write:
\begin{equation}\label{fullxsec}
\frac{d^2\sigma}{dx dy} = K(Q^2)\left[Y_+F_2(x,Q^2) - y^2 F_L(x,Q^2) \mp
  Y_- xF_3(x,Q^2)\right]\,,
\end{equation}
with $Y_{\pm} \equiv 1 \pm (1-y)^2$, where $x$, $Q^2$ and $y$ are the usual  DIS
variables and $K(Q^2)$ is
a kinematic factor different for neutral- and charged-current scattering.

Typically,  experimental measurements are given
in terms in of dimensionless reduced cross sections~\cite{Aaron:2012qi}, 
defined
as:
\begin{equation}\label{redxsec}
\widetilde{\sigma}(x,Q^2,y)\equiv 
\left\{ \begin{array}{ll}
\displaystyle \frac{1}{KY_+} \frac{d^2\sigma}{dx dy} =
F_2(x,Q^2) - \frac{y^2}{Y_+} F_L(x,Q^2) \mp \frac{Y_-}{Y_+} xF_3(x,Q^2) &\quad \mbox{for NC}\\
\\
\displaystyle \frac1{K} \frac{d^2\sigma}{dx dy} =
Y_+ F_2(x,Q^2) - y^2 F_L(x,Q^2) \mp Y_- xF_3(x,Q^2) &\quad \mbox{for CC}
\end{array}\right. \,.
\end{equation}
In {\tt APFEL} we have implemented the reduced cross
sections on top of the individual structure functions.
In addition, {\tt APFEL} provides separated predictions for light
and heavy structure functions: 
\begin{equation}\label{SFdecomposition}
F_i = F_i^{l} + F_i^{c} + F_i^{b} + F_i^{t}\quad\mbox{with}\quad i = 2,3,L\,,
\end{equation}
where $F_i^{l}$ is the contribution due to the light flavours (up,
down and strange), $F_i^{c}$ to the charm, $F_i^{b}$ to the bottom and
$F_i^{t}$ to the top.
Note that at $\mathcal{O}\lp \alpha_s^2\rp$ for neutral-current
and at $\mathcal{O}\lp \alpha_s\rp$ for charged-current, there can
be ambiguities in the definition of heavy quark structure functions,
in {\tt APFEL} we follow the conventions of Refs.~\cite{Forte:2010ta} 
and~\cite{Ball:2011mu} respectively. 

Structure functions are related to parton distributions
 by means of the convolution:
\begin{equation}
F_i(x,Q^2) \equiv \sum_{j=g,q,\overline{q}} C_i^j(x,Q^2) \otimes
q_j(x,Q^2)\,,
\end{equation}
where the coefficient functions $C_i^j$ can be computed in
perturbation theory and are usually given as power series in the strong
coupling $\alpha_s$.
The computation of the coefficient functions $C_i^j$ can be performed
in different mass schemes and there are basically two possibilities
where the heavy quarks are treated either as massive, usually
referred to as fixed-flavour-number (FFN) scheme, or as massless
partons, usually called zero-mass variable-flavour-number
(ZM-VFN) scheme.
The resulting calculations are more accurate in two different
and complementary regimes: the FFN scheme is more accurate
for values of $Q^2$ comparable to the mass of the heavy quark involved
in the calculation, while the ZM-VFN scheme 
is instead more accurate for values of
$Q^2$ much larger than the heavy-quark mass.

There exist different prescriptions to combine the FFN and the
ZM-VFN schemes in such a way to obtain accurate predictions over the
complete $Q^2$ range, which are referred to as General-Mass
Variable-Flavour-Number (GM-VFN) schemes\cite{Forte:2010ta,Aivazis:1993pi,Guzzi:2011ew,Thorne:2012az}. 
{\tt APFEL}
implements the FONLL scheme~\cite{Forte:2010ta}, but it also
provides the possibility to compute predictions in the FFN and in
the ZM-VFN scheme separately.

\section{Numerical techniques}
\label{sec:methods}

In this section we will present the numerical techniques
 that {\tt APFEL} uses to
solve the DGLAP evolution equations. Both QCD and QED DGLAP
evolution equations have the same formal structure, and thus the same
numerical techniques presented in this section apply to both of them.
In order to show the general strategy, here we will see how {\tt APFEL}
solves the QCD evolution equations but keeping in mind that the same
procedure applies to the QED ones as well.

The QCD DGLAP evolution equations can be written as:
\begin{equation}\label{dglap}
\mu^{2}\frac{\partial q_{i}(x,\mu)}{\partial \mu^2}=\int^{1}_{x}\frac{dy}y P_{ij}\left(\frac{x}{y},\alpha_{s}(\mu)\right)q_{j}(y,\mu)\,, 
\end{equation}
where $ P_{ij}\left(x,\alpha_{s}(\mu)\right)$ are the usual QCD
splitting functions up to some perturbative order in $\alpha_s$.
If we make the following definitions:
\begin{equation}
\begin{array}{rcl}
t &\equiv&\ln(\mu^{2})\,,\\
\tilde{q}(x,t)&\equiv&xq(x,\mu)\,,\\
\tilde{P}_{ij}(x,t)&\equiv&xP_{ij}(x,\alpha_{s}(\mu))\,,
\end{array}
\end{equation}
Eq.~(\ref{dglap}) becomes:
\begin{equation}\label{dglap2}
\frac{\partial \tilde{q}_{i}(x,t)}{\partial t}=\int^{1}_{x}\frac{dy}y \tilde{P}_{ij}\left(\frac{x}{y},t\right)\tilde{q}_{j}(y,t) \ .
\end{equation}
In order to numerically solve the above equation, we choose to express PDFs 
in terms of an interpolation basis over an $x$ grid with $N_x+1$ points. 
This way we can write:
\begin{equation}
\tilde{q}(y,t)=\sum^{N_{x}}_{\alpha=0}w_{\alpha}^{(k)}(y)\tilde{q}(x_{\alpha},t)\,,
\end{equation}
where $\{w_{\alpha}^{(k)}(y)\}$  is a set of interpolation functions
of degree $k$. In {\tt APFEL} we have chosen to use the Lagrange
interpolation method and therefore the interpolation functions read:
\begin{equation}\label{LagrangeFormula}
w_{\alpha}^{(k)}(x) = \sum_{j=0,j \leq \alpha}^{k}\theta(x-x_{\alpha-j})\theta(x_{\alpha-j+1}-x)\prod^{k}_{\delta=0,\delta\ne j}\left[\frac{x-x_{\alpha-j+\delta}}{x_{\alpha}-x_{\alpha-j+\delta}}\right]\,.
\end{equation}
Notice that Eq. (\ref{LagrangeFormula}) implies that:
\begin{equation}\label{nonzero}
w_{\alpha}^{(k)}(x) \neq 0 \quad\mbox{for}\quad x_{\alpha-k} < x < x_{\alpha+1} \,.
\end{equation}
Now we can rewrite Eq.~(\ref{dglap2}) as follows:
\begin{equation}\label{dglap3}
\frac{\partial \tilde{q}_{i}(x,t)}{\partial t}=\sum_{\alpha}\left[\int^{1}_{x}\frac{dy}y \tilde{P}_{ij}\left(\frac{x}{y},t\right)w_{\alpha}^{(k)}(y)\right]\tilde{q}_{j}(x_{\alpha},t) \, .
\end{equation}
In the particular case in which the $x$ variable in Eq. (\ref{dglap3}) coincides with one of the $x$-grid nodes, say
$x_\beta$, the evolution equations take the following discretized form:
\begin{equation}\label{dglap4}
\frac{\partial \tilde{q}_{i}(x_\beta,t)}{\partial t}=\sum_{\alpha}\underbrace{\left[\int^{1}_{x_\beta}\frac{dy}y \tilde{P}_{ij}\left(\frac{x_\beta}{y},t\right)w_{\alpha}^{(k)}(y)\right]}_{\Pi_{ij,\beta\alpha}(t)}\tilde{q}_{j}(x_{\alpha},t)\,.
\end{equation}
From Eq.~(\ref{nonzero}) follows the condition:
\begin{equation}\label{nonzero2}
\Pi_{ij,\beta\alpha}(t) \neq 0 \quad\mbox{for}\quad \beta \leq \alpha\,.
\end{equation}
In addition, the computation $\Pi_{ij,\beta\alpha}$ in  Eq.~(\ref{dglap4}) can be simplified to:
\begin{equation}\label{optimization}
\Pi_{ij,\beta\alpha}(t) = \int^{b}_{a}\frac{dy}y \tilde{P}_{ij}\left(\frac{x_\beta}{y},t\right)w_{\alpha}^{(k)}(y)\,,
\end{equation}
where the integration bounds are given by:
\begin{equation}
a \equiv \mbox{max}(x_\beta,x_{\alpha-k})\quad\mbox{and}\quad b \equiv \mbox{min}(1,x_{\alpha+1})\,.
\end{equation}
Alternatively, by means of a change of variable, the integral in
Eq. (\ref{optimization}) can be rearranged as follows:
\begin{equation}\label{optimization2}
\Pi_{ij,\beta\alpha}(t) = \int^{d}_{c}\frac{dy}y \tilde{P}_{ij}(y,t)w_{\alpha}\left(\frac{x_\beta}{y}\right)\,,
\end{equation}
where the new integration bounds are defined as:
\begin{equation}\label{bounds2}
c \equiv \mbox{max}(x_\beta,x_\beta/x_{\alpha+1}) \quad\mbox{and}\quad d \equiv \mbox{min}(1,x_\beta/x_{\alpha-k}) \,.
\end{equation}

One central aspect of the numerical methods used in
{\tt APFEL} is the use of an interpolation over a
logarithmically distributed $x$ grid.
In this case, the interpolation
coefficients in Eq.~(\ref{LagrangeFormula}) can be expressed as
\begin{equation}\label{LagrangeFormulaLog}
w_{\alpha}^{(k)}(x) = \sum_{j=0,j \leq \alpha}^{k}\theta(x-x_{\alpha-j})\theta(x_{\alpha-j+1}-x)\prod^{k}_{\delta=0,\delta\ne j}\left[\frac{\ln(x)-\ln(x_{\alpha-j+\delta})}{\ln(x_{\alpha})-\ln(x_{\alpha-j+\delta})}\right]\,.
\end{equation}
If in addition the $x$ grid is logarithmically distributed,
\textit{i.e.} such that
$\ln(x_{\beta})-\ln(x_{\alpha})=(\beta-\alpha)\Delta$, where the step
$\Delta$ is a constant, one has that the interpolating functions read:
\begin{equation}\label{LagrangeFormulaLog2}
w_{\alpha}^{(k)}(x) = \sum_{j=0,\,j \leq \alpha}^{k}\theta(x-x_{\alpha-j})\theta(x_{\alpha-j+1}-x)\prod^{k}_{\delta=0,\delta\ne j}\left[\frac{1}{\Delta} \ln\left(\frac{x}{x_\alpha}\right)\frac{1}{j-\delta}+1\right]\,,
\end{equation}
so that the dependence on $x$ of the interpolating function $w_\alpha^{(k)}(x)$ is
through the function $ \ln(x/x_\alpha)$ only. Therefore,
it can be shown that in Eq.~(\ref{optimization2})
$w_{\alpha}^{(k)}\left(x_\beta/y\right)$ depends only on the
combination $\lc \lp \beta-\alpha \rp \Delta-\ln y\rc$ and thus
$\Pi_{ij,\beta\alpha}$ depends only on the difference $(\beta-\alpha)$. 

One can use this information, together with the condition in
Eq.~(\ref{nonzero2}), to represent $\Pi_{ij,\beta\alpha}(t)$ as a
matrix, where $\beta$ is the row index and $\alpha$ the column index.
Such a representation of $\Pi_{ij,\beta\alpha}(t)$ reads:
\begin{equation}\label{MatrixRep}
\displaystyle \Pi_{ij,\beta\alpha}(t) = 
\begin{pmatrix}
a_0 &  a_1 & a_2 & \cdots & a_{N_x} \\
 0  & a_0 & a_1 & \cdots & a_{N_x-1} \\
 0  & 0   &  a_0 & \cdots & a_{N_x-2} \\
\vdots & \vdots & \vdots & \ddots & \vdots \\
 0  &   0  &   0 & \cdots & a_0 
\end{pmatrix}\,.
\end{equation}
Therefore, the knowledge of the first row of the matrix
$\Pi_{ij,\beta\alpha}(t)$ is enough to determine all the other entries.
This feature, which is based on the particular choice of the
interpolation procedure, leads to a more efficient computation of the
evolution operators since it reduces by a factor $N_x$ the number of
integrals to be computed.

After the presentation of the interpolation method, we turn to discuss the 
actual computation of the evolution operators. 
Any splitting function,
be it QED or QCD at any given perturbative order, has the following general structure:
\begin{equation}\label{SlittingFunctionDecomposition}
\tilde{P}_{ij}(x,t) = xP_{ij}^{R}(x,t) + \frac{xP_{ij}^{S}(x,t)}{(1-x)_+} + P_{ij}^{L}(t)x\delta(1-x)\,,
\end{equation}
where $P_{ij}^{R}(x,t)$ is the regular term,  $P_{ij}^{S}(x,t)$ is the
coefficient of the plus-distribution term, and $P_{ij}^{L}(t)$ is the
coefficient of the local term proportional to the delta functions.
It is useful to recall here that the general definition of
plus-distribution in the presence of arbitrary integration bounds is given by:
\begin{equation}\label{PlusPrescriptionDef}
\int_c^d dy \frac{f(y)}{(1-y)_+} = \int_c^d dy \frac{f(y) - f(1)\theta(d-1)}{1-y} + f(1)\ln(1-c)\theta(d-1)\,.
\end{equation}
Moreover, each of the functions $P_{ij}$ appearing in
Eq. (\ref{SlittingFunctionDecomposition}) has the usual perturbative
expansion that at N$^k$LO reads:
\begin{equation}
P_{ij}^{J}(x,t) = \sum_{n=0}^{k}a_s^{n+1}(t)P_{ij}^{J,(n)}(x),\quad\mbox{with}\quad J=R,S,L \,,
\label{eq:expansion}
\end{equation}
where we have defined $a_s\equiv \alpha_s/4\pi$.

Taking the above considerations into account and using the fact that
$w_{\alpha}^{(k)}(x_\beta)=\delta_{\beta\alpha}$, we can write the evolution operators
in terms of the various parts of the splitting functions as follows:
\begin{equation} \label{pertExp}
\begin{array}{c}
\displaystyle \Pi_{ij,\beta\alpha}(t) = \\
\\
\displaystyle \sum_{n=0}^{k} a_s^{n+1}(t) \bigg\{\int^{d}_{c}dy\left[{P}_{ij}^{R,(n)}(y)w_{\alpha}\left(\frac{x_\beta}{y}\right)+\frac{{P}_{ij}^{S,(n)}(y)}{1-y}\left(w_{\alpha}\left(\frac{x_\beta}{y}\right)-\delta_{\beta\alpha}\theta(d-1)\right)\right]\\
\\
\displaystyle +\left[{P}_{ij}^{S,(n)}(1)\ln(1-c)\theta(d-1)+{P}_{ij}^{L,(n)}\right]\delta_{\beta\alpha}\bigg\}\equiv \sum_{n=0}^{k} a_s^{n+1}(t) \Pi_{ij,\beta\alpha}^{(n)}\,,
\end{array}
\end{equation}
where the coefficients $\Pi_{ij,\beta\alpha}^{(n)}$ are independent of
the energy scale $t$,  and need to be evaluated a single time once the
$x$ interpolation grid and the evolution parameters have been defined.

Now we will show that Eq.~(\ref{pertExp}) respects the symmetry
conditions of Eq.~(\ref{MatrixRep}). We can distinguish two cases: 1)
$d < 1$ and  2) $d = 1$. In the case 1), due to the presence of the
Heaviside functions $\theta(d-1)$, Eq.~(\ref{pertExp}) reduces to:
\begin{equation}\label{pertExp2}
\Pi_{ij,\beta\alpha}^{(n)} = \int^{d}_{c}dy\left[{P}_{ij}^{R,(n)}(y)+\frac{{P}_{ij}^{S,(n)}(y)}{1-y}\right]w_{\alpha}\left(\frac{x_\beta}{y}\right) + {P}_{ij}^{L,(n)}\delta_{\beta\alpha}\,,
\end{equation}
which clearly follows Eq.~(\ref{MatrixRep}). In the case 2), instead, we have:
\begin{equation}\label{pertExp3}
\begin{array}{c}
\displaystyle \Pi_{ij,\beta\alpha}^{(n)} = \int^{1}_{c}dy\left[{P}_{ij}^{R,(n)}(y)w_{\alpha}\left(\frac{x_\beta}{y}\right)+\frac{{P}_{ij}^{S,(n)}(y)}{1-y}\left(w_{\alpha}\left(\frac{x_\beta}{y}\right)-\delta_{\beta\alpha}\right)\right]\\
\\
\displaystyle +\left[{P}_{ij}^{S,(n)}(1)\ln(1-c)+{P}_{ij}^{L,(n)}\right]\delta_{\beta\alpha}\,,
\end{array}
\end{equation}
and apparently, if $\alpha=\beta$, the term proportional to $\ln(1-c)$
could break the symmetry. However, from Eq.~(\ref{bounds2}), we know
that in this case:
\begin{equation}
c = \mbox{max}(x_\beta,x_\beta/x_{\beta+1}) = \frac{x_\beta}{x_{\beta+1}} \,,
\end{equation}
because $x_{\beta+1} \leq 1$. 
In addition, on a logarithmically distributed grid we have that
 $x_{\beta+1}=x_{\beta}\exp(\Delta)$. 
Therefore, it turns out that:
\begin{equation}
\ln(1-c) = \ln\left(1-\frac{x_{\beta}}{x_{\beta+1}}\right) = \ln[1 - \exp(-\Delta)]\,,
\end{equation}
which is a constant which does not depend on the indices $\alpha$ and $\beta$ and therefore
satisfies Eq.~(\ref{MatrixRep}).

At this point, the DGLAP equations imply that the discretized PDFs evolve between two scales $t$ and $t_0$
according to the following matrix equation:
\begin{equation}\label{EvolutionOperatorsDef}
\tilde{q}_{i}(x_\beta,t) = \sum_{\gamma,k} \Gamma_{ik,\beta\gamma}(t,t_0)\tilde{q}_{k}(x_\gamma,t_0) \ ,
\end{equation}
where it follows from Eq.~(\ref{dglap4}) that the evolution operators
are given by the solution of the system:
\begin{equation}\label{tosolve}
\left\{\begin{array}{l}
\displaystyle \frac{\partial  \Gamma_{ij,\alpha\beta}(t,t_0)}{\partial t}=\sum_{\gamma,k} \Pi_{ik,\alpha\gamma}(t)\Gamma_{kj,\gamma\beta}(t,t_0)\\
\\
\displaystyle \Gamma_{ij,\alpha\beta}(t_0,t_0)=\delta_{ij}\delta_{\alpha\beta}
\end{array}\right.
\end{equation}
Eq.~(\ref{tosolve}) is a set of coupled first order ordinary linear differential equations for the
evolution operators $\Gamma_{ij,\alpha\beta}(t,t_0)$.
In {\tt APFEL} Eq.~(\ref{tosolve}) is solved  using a fourth-order
adaptive step-size control Runge-Kutta (RK) algorithm.
Note that no interpolation in $t$ is involved, the solution of the
differential equations in $t$ is only limited by the precision of the RK method.
Once the evolved PDFs at the grid values $\tilde{q}_{i}(x_\beta,t)$ 
have been determined by means of the evolution operators in Eq.~(\ref{EvolutionOperatorsDef}), 
the value of these same PDFs for arbitrary values of $x$ will be computed
using again higher-order interpolation.

A final consideration concerning the choice of interpolating grid in
$x$ is needed.
As is well known, an accurate solution of the  DGLAP equations
requires a denser grid at large $x$, where PDFs have more structure than at
small-$x$.
In {\tt APFEL} it is not possible to use an $x$-grid with variable
spacing that allows to have a denser grid at large $x$ and at the same time to maintain
the symmetry that allows to substantially reduce the number of integrals to be
evaluated, see Eq.~(\ref{MatrixRep}).
In fact, a logarithmically distributed $x$ grid necessarily
leads to a looser grid in the large-$x$ region, thus potentially
degrading the evolution accuracy there.
To overcome this problem, {\tt APFEL} implements the possibility of using
different interpolation grids according to the value of $x$ in which
PDFs need to be evaluated.

The basic idea is the following. The evolution of a given set of PDFs
from the initial condition at the scale $\mu_0$ up to some other scale
$\mu$ is determined by the convolution between the evolution operators
and the boundary conditions, which implies performing and integral
between $x$ and one.
This convolution, when discretized on an interpolation $x$ grid,
corresponds to Eq.~(\ref{EvolutionOperatorsDef}).
It is clear that such operation will use only those $x_{\beta}$ nodes of
the interpolation grid that fall in the range between $x$ and one. 

Therefore, the computation of the PDF evolution in the large-$x$
region using a logarithmically spaced interpolation grid with a small value of $x_{\rm min}$ will
be certainly inefficient, since the convolution would use only a small
number of points
in the large-$x$ region such that $x_{\beta}\le x \le 1$, discarding those with  $x< x_{\beta}$.
In order to avoid this problem and simultaneously achieve a good
accuracy and performance over the whole range in $x$, 
{\tt APFEL} gives the possibility to use different interpolating
grids, each with a different value of $x_{\rm min}$, interpolation degree and number of points.
Then, to compute the evolution of the PDFs for the point $x$, the program
will automatically select the grid with the largest value of $x_{\rm
  min}$ compatible with the condition $x_{\rm min}\le x$. 
In the user's manual Sect.~\ref{sec-manual} we will describe the functions that
allow to select this option and achieve a good accuracy in the complete
$x$-range, while keeping always log-spaced interpolating grids.

The use of $n\ge 2$ subgrids increases slightly the time taken by initialization phase,
since more evolution operators need to be precomputed, and also the actual
evolution is somewhat slower than in the case with a single grid ($n=1$), 
with the important trade-off of a much
more accurate result in the large-$x$ region. 
As default settings, {\tt APFEL} uses $n=3$ interpolation
grids, with interpolation order $3,5$ and $5$, number of points $N_x=80,50$ and
$40$ and $x_{\rm min}=10^{-5},0.1$ and $0.8$ respectively.
These are the settings that have been used in all the benchmark comparisons
with other codes that we will discuss in Sect.~\ref{sec-benchmarks}.

\section{{\tt APFEL} library documentation}\label{sec-manual}

In this section we present the user manual for the  {\tt APFEL} library.
Written in {\scshape Fortran 77},
all the functionalities can also be accessed via the  {\tt C/C++} and
 {\tt Python} interfaces. 
For simplicity, we will restrict ourselves to the
description of the {\tt C/C++} interface, but the usage of
the {\scshape Fortran 77} and  {\tt Python} interfaces is
very similar and examples of their use are provided 
in the {\tt examples} folder of the {\tt APFEL} source code.
First of all, we will discuss how to install {\tt APFEL}  and how to execute the basic
example programs. After that, we will 
list the various customization options that can
be accessed by the user for both the PDF evolution and the DIS structure functions modules.
%
%
Finally, we will describe how to intall the {\tt APFEL} Graphical
User Interface (GUI), giving some ebasic examples on ho to use the associated plotting modules.

\subsection{Installation and basic execution}

The  {\tt APFEL} library is available from its {\tt HepForge} website:
\begin{center}
{\bf \url{http://apfel.hepforge.org/}~}
\end{center}
and it can also be accessed directly from the svn repository, both
for the development trunk:
\begin{center}
\tt svn checkout http://apfel.hepforge.org/svn/trunk apfel
\end{center}
as well as for the current stable release, in the case of v2.0.0
for instance one has:
\begin{center}
\tt svn checkout http://apfel.hepforge.org/svn/tags/2.0.0 apfel-2.0.0
\end{center}
The installation of the {\tt APFEL} library  can be easily  performed 
using the standard {\tt autotools} sequence:
\begin{lstlisting}
   ./configure
   make
   make install
\end{lstlisting}
which automatically installs {\tt APFEL} in  {\tt
  /usr/local/}. 
Note that the {\tt APFEL} library requires an installation of the {\tt LHAPDF} PDF library.\footnote{The
current release of {\tt APFEL} assumes that {\tt LHAPDF5.9.0} or
a more recent version has been previously installed.
}
To use a different installation path, one simply needs to
use the option:
\begin{lstlisting}
   ./configure --prefix=/path/to/the/installation/folder
\end{lstlisting}
In this case, the {\tt APFEL} installation path should be included
to the environmental variable {\tt LD\_LIBRARY\_PATH}.
This can be done adding to the local {\tt .bashrc} file (or {\tt
  .profile} file on Mac) the string:
\begin{lstlisting}
   export LD_LIBRARY_PATH=$LD_LIBRARY_PATH:/path/to/the/installation/folder/lib
\end{lstlisting}

Once {\tt APFEL} has been properly compiled and installed, the user
has at her/his disposal a set of routines that can be called from a
main program. In the installation {\tt bin} directory there is the
{\tt apfel-config} script, useful to determine the compiler flags in
custom {\tt makefiles}, together with a shell script {\tt apfel} which
starts an interactive console session of {\tt APFEL} providing an
immediate instrument to use the library without coding.
In the following
we will illustrate these functionalities and how
they can be accessed by the user.
The basic usage of {\tt APFEL}  requires
only two steps to have the complete set of evolved PDFs.
The first step is the initialization of {\tt APFEL} through the call
of the following routine:
\begin{lstlisting}
   InitializeAPFEL
\end{lstlisting}
This will precompute all the needed evolution operators that enter the
discretized DGLAP equation, as discussed in Sect.~\ref{sec:methods}. 
Let us recall that once the general settings of the
evolution have been defined (perturbative order, heavy quark
masses, reference value of $\alpha_s$, and so on), 
the initialization needs to be performed only once, irrespective of
the scales that are used in the PDF evolution.
The second step consists in performing the actual PDF evolution between the
initial scale {\tt Q0} and the final scale {\tt Q} (in GeV). This can
be achieved using the routine:
\begin{lstlisting}
   EvolveAPFEL(Q0,Q)
\end{lstlisting}
With this routine {\tt APFEL} numerically solves the 
discretized DGLAP equations in $t$ using the evolution operators
 precomputed in the initialization step.
%
%
Now the user can access the evolved PDFs at the scale {\tt Q}
via the use of the functions:
\begin{lstlisting}
   xPDF(i,x)
   xgamma(x)
\end{lstlisting}
where the real variable {\tt x} is the desired value of Bjorken-$x$
while the integer variable {\tt i} in the function {\tt xPDF}, which runs from $-$6 to
6, corresponds to quark flavor index according to the same convention
used in the {\tt LHAPDF} library, that is:
\begin{table}[h]
\centering
\begin{tabular}{rcccccccccccccc}
{\tt i}~:  & $-$6 &$-$5 &$-$4&$-$3&$-$2&$-$1&\;0\;&\;1\;&\;2\:&\;3\;&\;4\;&\;5\;&\;6\;\\ 
{\tt xPDF}~: &  $\bar{t}$&$\bar{b}$&$\bar{c}$&$\bar{s}$&$\bar{u}$&$\bar{d}$&
$g$&$d$&$u$&$s$&$c$&$b$&$t$ \\
\end{tabular}
\end{table}

In {\tt APFEL} we have explicitly separated the access to the quark
and gluon PDFs (via {\tt xPDF}) and from that to the photon PDF
(via {\tt xgamma}). Notice that the functions {\tt xPDF} and {\tt
  xgamma} return $x$ times the PDFs (the momentum fractions).

In addition to the PDF values, the user can also access the integer Mellin moments of
the PDFs,\footnote{We follow the standard definition of
Mellin moments:
\be
{\tt NPDF(i,N)}\equiv \int_0^1 dx~x^{{\tt N-2}}\, {\tt xPDF(i,x)} \, .
\ee
} using the routines:
\begin{lstlisting}
   NPDF(i,N)
   Ngamma(N)
\end{lstlisting}
which are useful for instance to evaluate the momentum 
and valence sum rules at the scale {\tt Q}, using {\tt N=2} and
{\tt N=1} respectively.
Finally, two functions return the value of the QCD coupling $\alpha_s$ and of the QED
coupling $\alpha$ using the same settings used for the PDF evolution, these are:
\begin{lstlisting}
   AlphaQCD(Q)
   AlphaQED(Q)
\end{lstlisting}
In {\tt APFEL} we use the exact numerical solution of the QCD beta function
equations
using Runge-Kutta methods, while for the QED coupling the analytical
leading-order solution is used.

The basic information above is enough to write a simple and yet complete
program to perform PDF evolution using {\tt APFEL}.
As an illustration, a {\tt C/C++} program that computes and tabulates
PDFs to be compared with the Les Houches PDF benchmark evolution 
tables~\cite{lh2,Dittmar:2005ed} would be the following:
\begin{lstlisting}
#include <iostream>
#include <iomanip>
#include <cmath>
#include "APFEL/APFEL.h"
using namespace std;

int main()
{
 // Define grid in x
  double xlha[11] = {1e-7, 1e-6, 1e-5, 1e-4, 1e-3, 1e-2, 
		1e-1, 3e-1, 5e-1, 7e-1, 9e-1};
  
 // Precomputes evolution operators on the grids nodes
  APFEL::InitializeAPFEL();

  // Set initial and final evolution scales
  double Q02, Q2,
  cout << "Enter initial and final scales in GeV2" << endl;
  cin >> Q02 >> Q2;

  // Perform evolution
  double Q0 = sqrt(Q02);
  double Q  = sqrt(Q2);
  APFEL::EvolveAPFEL(Q0,Q);

  cout << scientific << setprecision(5) << endl;
  
  cout << "alpha_QCD(mu2F) = " << APFEL::AlphaQCD(Q) << endl;
  cout << "alpha_QED(mu2F) = " << APFEL::AlphaQED(Q) << endl;
  cout << endl;

  cout << "   x   " 
       << setw(11) << "   u-ubar   " 
       << setw(11) << "   d-dbar   " 
       << setw(11) << " 2(ubr+dbr) " 
       << setw(11) << "   c+cbar   " 
       << setw(11) << "   gluon    " 
       << setw(11) << "   photon   " << endl;

  cout << scientific;

  // Tabulate PDFs for the LHA x values
  for (int i = 0; i < 11; i++)
    cout << xlha[i] << "\t"  
	 << APFEL::xPDF(2,xlha[i]) - APFEL::xPDF(-2,xlha[i]) << "\t"
	 << APFEL::xPDF(1,xlha[i]) - APFEL::xPDF(-1,xlha[i]) << "\t"
	 << 2*(APFEL::xPDF(-1,xlha[i]) + APFEL::xPDF(-2,xlha[i])) << "\t"
	 << APFEL::xPDF(4,xlha[i]) + APFEL::xPDF(-4,xlha[i]) << "\t"
	 << APFEL::xPDF(0,xlha[i]) << "\t"
	 << APFEL::xgamma(xlha[i]) << "\t"
	 << endl;

  return 0;
}
\end{lstlisting}
This example code uses the default settings of {\tt APFEL} for the
evolution parameters such as perturbative order, heavy quark masses, values of the couplings etc.
Such default settings correspond to those of the Les Houches benchmark at NNLO~\cite{lh2,Dittmar:2005ed}.
In the following we will discuss how the user can choose her/his own settings
for the PDF evolution in {\tt APFEL}.

\subsection{Customization of the PDF evolution}\label{EvolCustom}

The customization of the PDF evolution with {\tt APFEL} can be achieved using a number of dedicated 
routines, to be called before the initialization stage, that is before
calling {\tt InitializeAPFEL}.
These routines are:
\begin{itemize}

\item {\tt SetTheory(Theory)}: this routine defines the theory to be used
  for the PDF evolution. Using the notation for the various options for
  combining QCD and QED evolution introduced in
  Sect.~\ref{sec-theory}, the string variable {\tt Theory} can take
  the following values:
\begin{itemize}
\item {\tt "QCD"} for pure QCD,
\item {\tt "QED"} for pure QED,
\item {\tt "QCEDP"} for QCD$\otimes$QED in parallel,
\item {\tt "QCEDS"} for QCD$\otimes$QED in series,
\item {\tt "QECDP"} for QED$\otimes$QCD in parallel,
\item {\tt "QECDS"} for QED$\otimes$QCD in series,
\item {\tt "QavDP"} for the averaged solution in parallel,
\item {\tt "QavDS"} for the averaged solution in series.
\end{itemize}
Let us recall that all the options for the solution of the combined
QCD$\otimes$QED evolution equations are equivalent up to subleading $\mathcal{O}\lp \alpha \alpha_s\rp$
terms.

\item {\tt SetPerturbativeOrder(pt)}: this routine sets the
  perturbative order of the QCD evolution. 
The integer variable {\tt pt} can
  take the values 0, 1 or 2 corresponding to LO, NLO and NNLO
  evolution respectively.
The QED evolution, when activated, is always LO.

\item {\tt SetAlphaQCDRef(alphasref,Qref)}: this routine sets the
 reference value of the QCD coupling $\alpha_s$, {\tt alphasref},
 at the reference scale, {\tt Qref} in GeV. 

\item {\tt SetAlphaQEDRef(alpharef,Qref)}: same as {\tt
    SetAlphaQCDRef} but for the QED coupling $\alpha$.

\item {\tt SetPoleMasses(mc,mb,mt)}: this routine sets the values for
  the heavy quark masses in the pole mass scheme. The real variables {\tt mc},
 {\tt mb} and {\tt mt} correspond to the numerical values in GeV of the pole heavy quark masses
 $m_c$, $m_b$ and $m_t$. Calling this routine also determines that
 pole heavy quark masses are used as thresholds for the VFN scheme PDF evolution.
 
\item {\tt SetMSbarMasses(mc,mb,mt)}: this routine sets the values for
  the heavy quark masses in the $\MSbar$  scheme. Here the real variables {\tt mc},
 {\tt mb} and {\tt mt} correspond to the numerical values in GeV of the 
renormalization-group-invariant (RGI) heavy quark masses
$m_c(m_c)$, $m_b(m_b)$ and $m_t(m_t)$. Calling this routine also
determines that $\MSbar$ heavy quark masses are used as thresholds for
the VFN scheme PDF evolution.

\item {\tt SetRenFacRatio(Ratio)}: this routine sets the ratio between
  renormalization and factorization scales. The real variable {\tt Ratio}
  corresponds to the ratio $\mu_R / \mu_F$.
The default choice in {\tt APFEL} is {\tt Ratio=1}.
The modifications of the solutions of the DGLAP evolution equations
 for $\mu_R \ne \mu_F$ are discussed
for instance in Sect. 2.2 of Ref. \cite{pegasus}.

\item {\tt SetVFNS}: this routine determines that the
  variable-flavor-number scheme is used  for the PDF evolution.

\item {\tt SetFFNS(NF)}: this routine determines that the
  fixed-flavor-number scheme is used  for the PDF evolution. The
  integer variable {\tt NF} corresponds to the number
  of active quark flavor and can then take any values between 3 and 6.

\item {\tt SetMaxFlavourAlpha(NF)}: this routine sets the maximum
  number of active flavors that enter the QCD and QED beta functions
  for the $\alpha_s$ and $\alpha$
  running. The integer variable {\tt NF} can then take any value
  between 3 and 6.

\item {\tt SetMaxFlavourPDFs(NF)}: this routine sets the maximum
  number of active flavors that can contribute to the PDF
  evolution. The integer variable {\tt NF} can then take any value
  between 3 and 6.

\item {\tt SetPDFSet(name)}: this routine defines the PDF set to be
  evolved from the initial to the final scale. The string variable
  {\tt name} can take the value {\tt "ToyLH"}, corresponding to the toy
  PDF model used in the Les Houches PDF benchmarks~\cite{Giele:2002hx}, 
or the name (including
  the {\tt LHgrid} extension) of any PDF set available from the {\tt
    LHAPDF} library. 

There is yet a third option, {\tt name="private"},  which can be easily modified by the user (in {\tt src/toyLHPDFs.f}) if
new PDF boundary conditions not covered by the two other options
are required.
Note that each time a new {\tt private} parametrization is coded,
the library needs to be complied and installed again.
 By default, the routine {\tt
    private} is set to the  toy parametrization adopted in the 
older benchmark study of Ref.~\cite{Blumlein:1996rp}.

\item {\tt SetReplica(irep)}: this routine selects the replica (for a
  Monte Carlo PDF set) or the specific eigenvector  (for Hessian PDF
  sets) of the  PDF set defined above to be evolved with {\tt APFEL}.
  The integer variable {\tt irep} can then take any value included
  between 0 (the central PDFs)  and the maximum number of PDF members
  contained in the selected PDF set.

\item {\tt SetQLimits(Qmin,Qmax)}: this routine sets the scale
  bounds between which the evolution can be performed. 
The real
  variables {\tt Qmin} and {\tt Qmax} correspond to the numerical
  values of the lower and upper bounds (in GeV).
 On top of making sure
  that PDF evolution is performed only in the physical range, this
  option also allows to reduce the initialization time, for instance
  if {\tt Qmax} is below some heavy quark thresholds, reducing the number of
  evolution operators to be precomputed will be smaller.

\item {\tt SetNumberOfGrids(n)}: this routine sets the number of
  $x$-space interpolation grids that will be used for the
  evolution. The integer variable {\tt n} can be any positive integer number.

\item {\tt SetGridParameters(i,np,deg,xmin)}: this routine sets the
  parameters of the {\tt i}-th $x$-space interpolation grid. 
The
  integer variable {\tt i} must be between 1 and {\tt n}, where
the latter has been defined in
  {\tt SetNumberOfGrids(n)}.
 The integer variable {\tt np} corresponds
  to the number of (logarithmically distributed) points of the grid,
  the integer variable {\tt deg} corresponds to the interpolation
  degree and the real variable {\tt xmin} corresponds to the lower
  bound of the grid. The upper bound is always taken to be equal to one.
\end{itemize}

As an illustration, if the user wants to perform the QCD evolution at NLO
instead of  the default NNLO, she/he needs to add to the code above,
before the initialization routine {\tt InitializeAPFEL}, a call to the
corresponding function, that is:
\begin{lstlisting}
 APFEL::SetPerturbativeOrder(1);
\end{lstlisting}
or if the user wants to use as a boundary condition for the PDF
evolution a particular set available through the {\tt LHAPDF} interface, say
{\tt NNPDF23\_nlo\_as\_0118\_qed.LHgrid}, she/he needs
to call before the initialization the following function:
\begin{lstlisting}
 APFEL::SetPDFSet("NNPDF23_nlo_as_0118_qed.LHgrid");
\end{lstlisting}
By default, {\tt APFEL} will use the central replica of
the selected PDF set. 
Varying any other setting is similar, various example programs
have been collected in the {\tt examples} folder in the {\tt APFEL} source
folder.

When modifying the default settings, particular care 
must be taken with the number of interpolation grids, the number
of points in each grid and the order of the interpolation.
The  default
settings in {\tt APFEL} use three grids whose ranges and
number of points have been tuned to give accurate and
fast results over a wide range of $x$, as discussed
in Sect.~\ref{sec:methods}.
If the default parameters are modified, the user should check
that the accuracy is still good enough, by comparing for instance
with another run of {\tt APFEL} with the default interpolation
parameters.

The folder {\tt examples} in the {\tt APFEL} source directory contains
several examples that further illustrate the functionalities of
the code, and that can be used by the user as a starting point
towards a program that suits her/his particular physics needs.
All these examples are available in the three possible interfaces to
{\tt APFEL}: {\scshape Fortran 77}, {\tt C/C++} and {\tt Python}.

\subsection{Computation of the DIS observables}

Now we turn to the description of the module that computes the DIS
neutral- and charged-current observables. 
%
This module can be either used together with the PDF
evolution provided by {\tt APFEL} or directly interfaced to the
 {\tt LHAPDF} library.

The computation of  DIS  structure functions
is provided by a single routine,
 {\tt DIS\_xsec}, which takes a set of input parameters needed to
 specify the computation to be performed.
The usage of the {\tt DIS\_xsec} routine is the following:
\begin{lstlisting}
APFEL::DIS_xsec(x,q0,q,y,proc,scheme,pto,pdfset,irep,target,proj,F2,F3,FL,sigma);
\end{lstlisting}
where the \textit{input} parameters are:
\begin{itemize}

\item the real variable {\tt x}: the value of Bjorken $x$,

\item the real variable {\tt q0}: the value of the initial scale (in
  GeV) used in the PDF evolution (this input is ignored if the {\tt LHAPDF} evolution
is used),

\item the real variable {\tt q}: the value of the scale (in GeV) where the DIS
 observables are to be computed,

\item the real variable {\tt y}: the value of the inelasticity,

\item the string variable {\tt proc}: it can take the values {\tt "EM"} for the
  purely electromagnetic DIS observables (photon-only exchange), {\tt "NC"}
  for neutral-current observables and {\tt CC"} for charged-current
  observables,

\item the string variable {\tt scheme}: it can take the values {\tt "FONLL"} for FONLL, {\tt "FFNS"}
  for the FFN scheme and {\tt ZMVN"} for the ZM-VFN scheme,

\item the integer variable {\tt pto},: it can take the values {\tt
    0}, {\tt 1} or {\tt 2} corresponding to LO, NLO and NNLO,
  respectively. Notice that choosing {\tt
    pto=1} with {\tt scheme="FONLL"} implies using the
  FONLL-A scheme, while choosing {\tt pto=2} with {\tt
    scheme="FONLL"} leads to using the
  FONLL-C scheme. The implementation of
 the FONLL-B scheme~\cite{Forte:2010ta} is postponed to a future release of the program.

\item The string variable {\tt pdfset}: it can take any of the PDF sets
  available in {\tt LHAPDF} (including the {\tt .LHgrid} extension). This way {\tt APFEL} will use
  the selected PDF set to compute the DIS observables using the {\tt
    LHAPDF} evolution rather than the internal one. As an alternative, the user can choose {\tt
    pdfset="APFEL"}. This way the DIS observables will be computed
  using the evolution provided by {\tt APFEL} between the scales {\tt q0} and {\tt
    q}. For the setting of the PDF evolution, the user can refer to
  Sect.~\ref{EvolCustom}.

\item The integer value {\tt irep}: it specifies the member of the PDF set to be used,

\item the string variable {\tt target}: it takes the value {\tt
    "PROTON"} in case the target is a proton, {\tt "NEUTRON"} in case
  the target is a neutron (assuming isospin symmetry) or {\tt
    "ISOSCALAR"} if the target is an isoscalar, $e.g.$ a
  deuteron (also this option assumes isospin asymmetry). There is a
  further option, which is {\tt target="IRON"}, which uses the cross-section
  definition used in the NuTeV experiment which is on iron
  nuclei~\cite{Goncharov:2001qe}.

\item The string variable {\tt proj}: it takes the values {\tt
    "ELECTRON"} and {\tt "POSITRON"} if {\tt proc="EM", "NC"},
  in case the projectile is either an electron or a positron. If
  instead {\tt proc="CC"}, the variable {\tt proj} can also take the
  values {\tt "NEUTRINO"} or {\tt "ANTINEUTRINO"} with obvious meaning.
\end{itemize}

Once all these input parameters have been specified, the \textit{output}
array variables are {\tt F2, F3, FL} and {\tt sigma}. Each of them has
5 entries corresponding to light, charm, bottom, top and total
components of the corresponding quantity, as described in
Sect.~\ref{sec-dis}.
The user should be careful because in the {\tt Fortran} interface the arrays
are numbered from 3 to 7 ($e.g.$ {\tt F2(3)} = $F_2^l$, {\tt F2(4)} =
$F_2^c$, {\tt F2(5)} = $F_2^b$, {\tt F2(6)} = $F_2^t$, {\tt F2(7)} =
$F_2^p$), while in the C++ version the arrays run from 0 to 4 ($e.g.$ {\tt F2(0)} = $F_2^l$, {\tt F2(1)} =
$F_2^c$, {\tt F2(2)} = $F_2^b$, {\tt F2(3)} = $F_2^t$, {\tt F2(4)} =
$F_2^p$).
As in the case of the PDF evolution, the user will find an example on how to use the DIS module in
{\tt example} folder.


\subsection{The Graphical User Interface}
\label{sec-gui}

The {\tt APFEL} package provides also a Graphical User Interface ({\tt
  GUI}) available in the {\tt apfel/apfelGUI} folder
of the source code. The {\tt GUI} provides a simple and fast way to
access almost all the features implemented in {\tt APFEL} without
the need to write any code.
To use the {\tt GUI}, the system
requirements are {\tt Qt4}, a cross-platform application and UI
framework, {\tt LHAPDF} for the PDF manipulation, and {\tt ROOT} for the
plotting resources.

The application is compiled by using the standard {\tt qmake} commands:
\begin{lstlisting}
   cd apfelGUI/
   qmake
   make
\end{lstlisting}
This will produce the executable {\tt apfelgui} that is located in the
{\tt apfelGUI} folder itself, unless the user is running on a Mac OS X
operating system equipped with the Xcode. In this case the executable
should be found in the {\tt apfelGUI/apfelgui.app/Contents/MacOS/apfelgui}
folder.\footnote{In case the default language of the operating system
  is different from english, we recommend to use the following command to run the
  {\tt GUI}:
$$
\mbox{\tt LC\_ALL=C ./apfelgui.app/Contents/MacOS/apfelgui}
$$
This ensures that the floating-point numbers are treated consistently.}

\begin{figure}[h]
\centering
\includegraphics[scale=0.35]{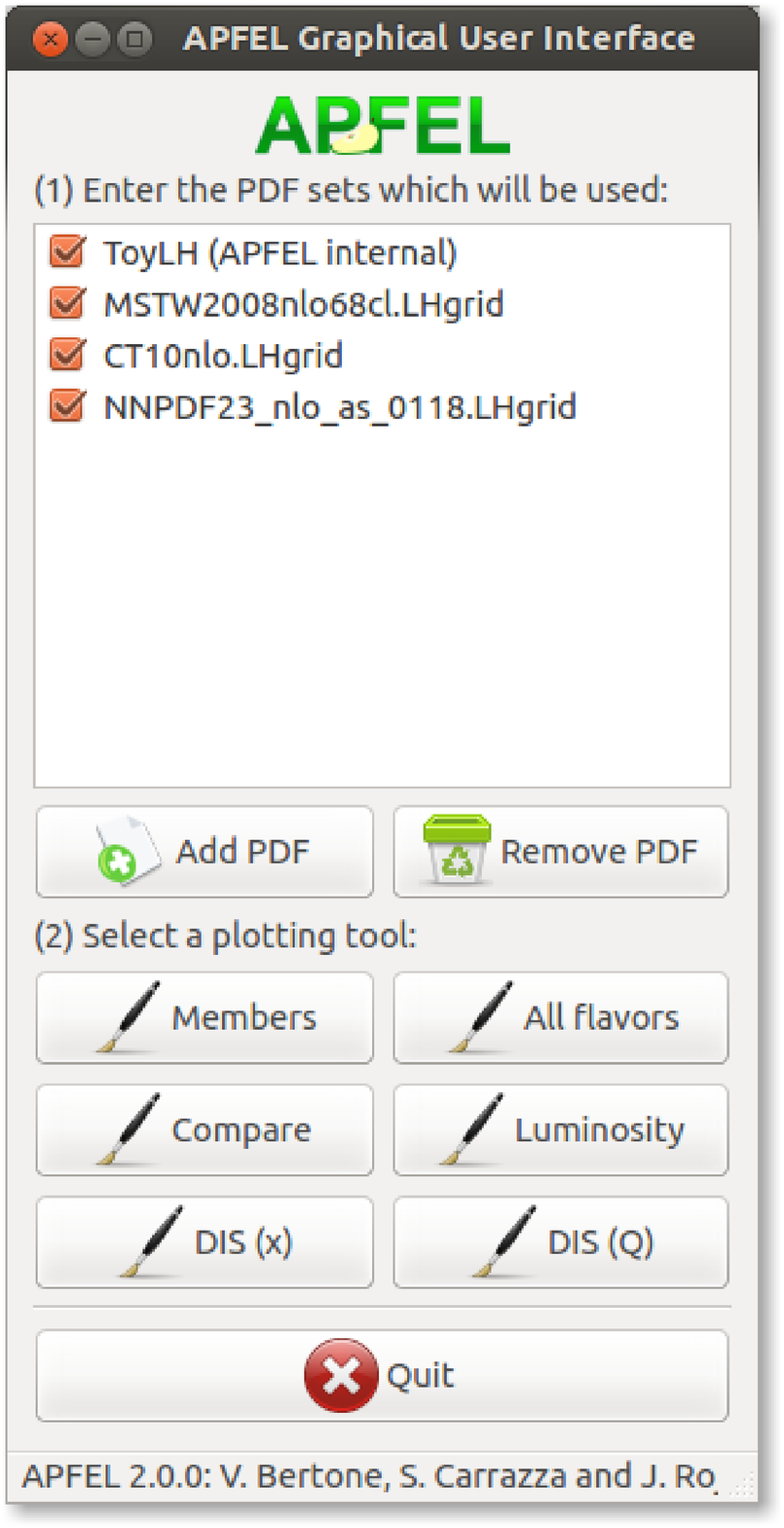}
\includegraphics[scale=0.35]{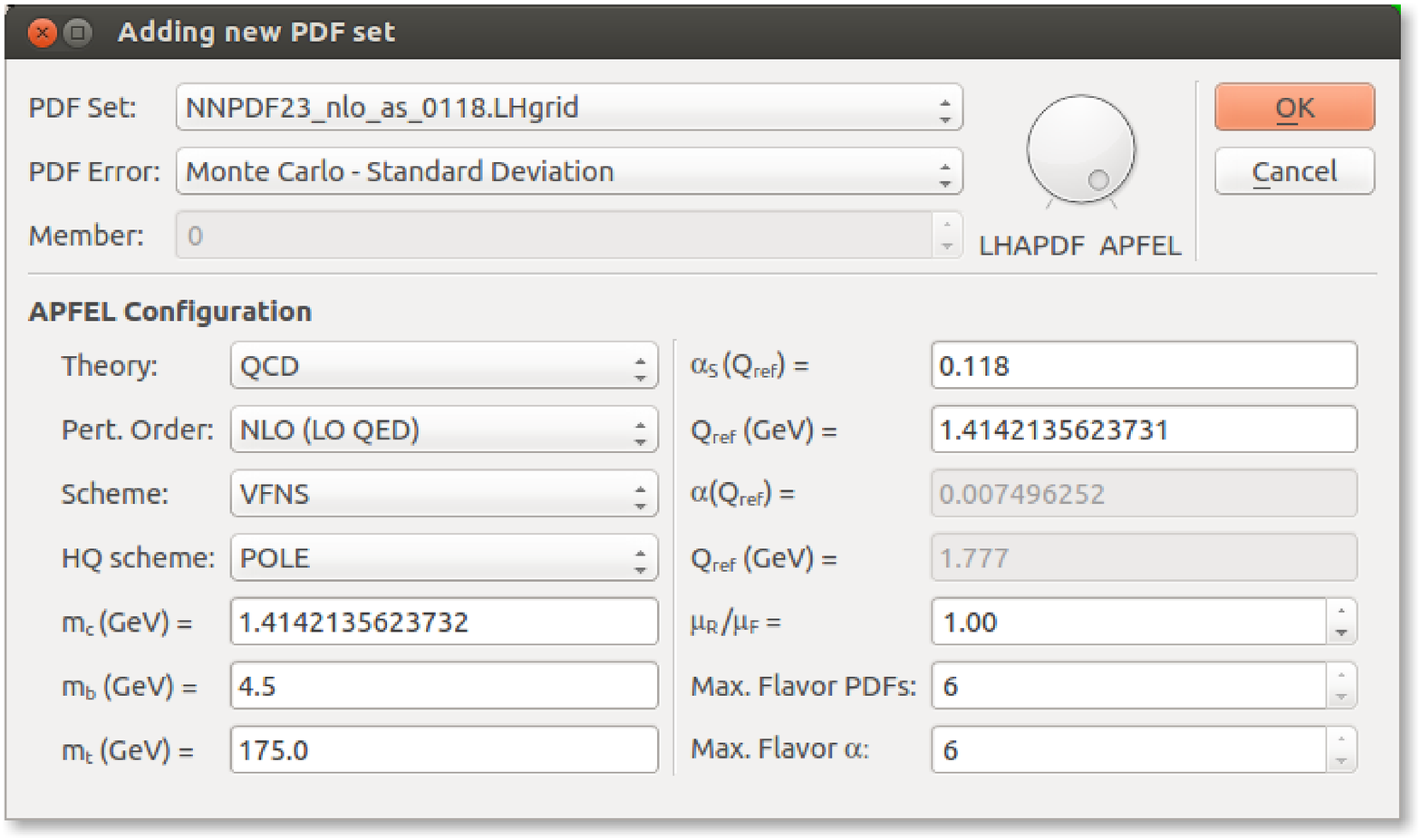}
\caption{Snapshot of the main window on the left, and the PDF setup
  dialog on the right.}
\label{fig:gui}
\end{figure}

While the use of the {\tt GUI} should be self-explanatory,
let us briefly present some of its main ingredients.
In Fig.~\ref{fig:gui} we show a snapshot of the main window and the
PDF dialog where the user can choose the PDFs to be used and
select the desired plotting tool. The program handles the two most
common method for quantifying PDF uncertainties, the
 Monte Carlo and the Hessian methods. Alternatively, the user can
 select a specific PDF member/replica to perform the computation. This
 last option is particularly interesting when considering PDF sets for $\alpha_s$
variation. 
Once the PDF set has been selected, the PDF dialog allows the user to choose between two main
possibilities: the use {\tt LHAPDF} embedded evolution or
the {\tt APFEL} internal evolution taking the PDF input at some initial
scale to be specified in the command dashboard of each plotting tool.

The {\tt APFEL GUI}, starting from version 2.0.0,
 contains the following plotting tools:
\begin{itemize}
\item the ``{\tt Members}'' tool: it plots all the members of a PDF set
  for a single parton flavor at a user-defined energy scale.
\item The ``{\tt All flavors}'' tool: each PDF flavor is plotted together
  in the same canvas. We also provide the possibility to scale PDF
  flavors by a predetermined factor.
\item The ``{\tt Compare}'' tool: it compares the same flavor of multiple PDF
  sets and the respective uncertainties.
\item The ``{\tt Luminosity}'' tool: it performs the computation of parton
  luminosities~\cite{Ball:2010de} normalized to a reference PDF set.
\item The ``{\tt DIS observables}'' tool: it computes the DIS observables as
  functions of $x$ or $Q$ for different heavy quark schemes.
\end{itemize}

\begin{figure}[h]
\centering
\includegraphics[scale=0.35]{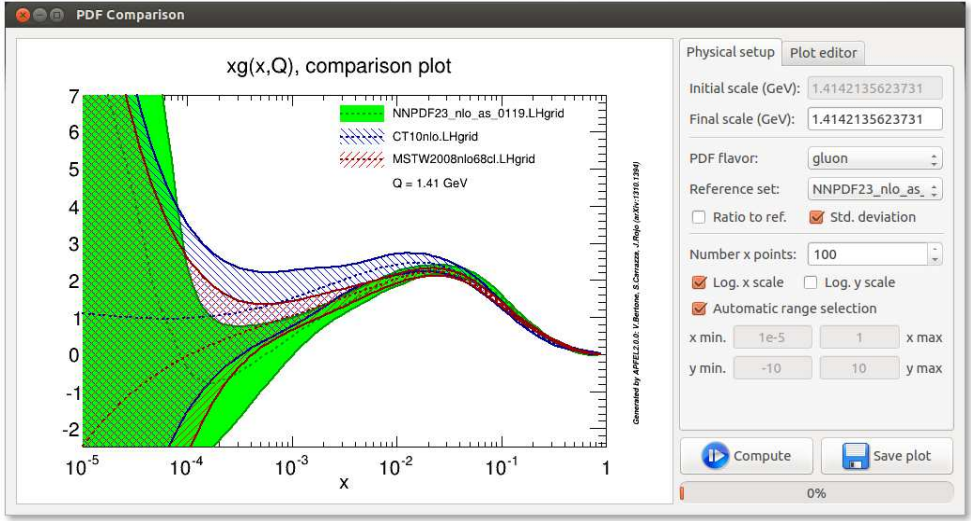}
\caption{Snapshot of the ``{\tt Compare}'' tool available in the {\tt APFEL}
  Graphical User Interface.}
\label{fig:gui2}
\end{figure}

For all the various 
tools, there are several options for the graphical customization,
like setting the axis ranges and axis titles. In Fig.~\ref{fig:gui2} an example
of the ``{\tt Compare}'' tool is showed. 
{\tt APFEL} also provides the possibility to
save plots and the associated underlying data in multiple formats. 
All the results provided by the
{\tt GUI} for PDFs and parton luminosities from
different PDF sets have been verified against the corresponding
results from the PDF benchmarking
exercise of Ref.~\cite{Ball:2012wy}.

\section{Validation and benchmarking}

\label{sec-benchmarks}

Having presented the methodology and the numerical techniques
that {\tt APFEL} adopts to solve the coupled QCD$\otimes$QED
equations, and provided the relevant user
documentation, we now 
turn to compare {\tt APFEL} with other publicly available codes. 
First of all, we perform a detailed benchmarking of {\tt APFEL}
against {\tt HOPPET} finding good agreement for the
QCD evolution up to NNLO, both with pole and $\MSbar$ heavy
quark masses.
Then we turn to the validation of the combined QCD$\otimes$QED
evolution, and we compare the predictions of {\tt APFEL}, using various options for the
solution of the coupled evolution equations, with the
{\tt partonevolution} code, with the internal MRST04QED 
evolution and with the NNPDF internal code {\tt FastKernel}.

In this section we also provide a study of the consistency of the different 
methods for the solution of the coupled  QCD$\otimes$QED
evolution equations, and show that when the {\tt QECD} and {\tt QCED}
solutions are constructed iteratively in small steps in $Q^2$
so as to avoid introducing potentially large logarithms, they
reduce numerically to the {\tt QavD} solution.
Finally, we provide results for the benchmarking of the
DIS structure function module of {\tt APFEL}.

\subsection{QCD evolution}\label{APFELvsHOPPET}

\begin{figure}[t]
\centering
\includegraphics[scale=0.4]{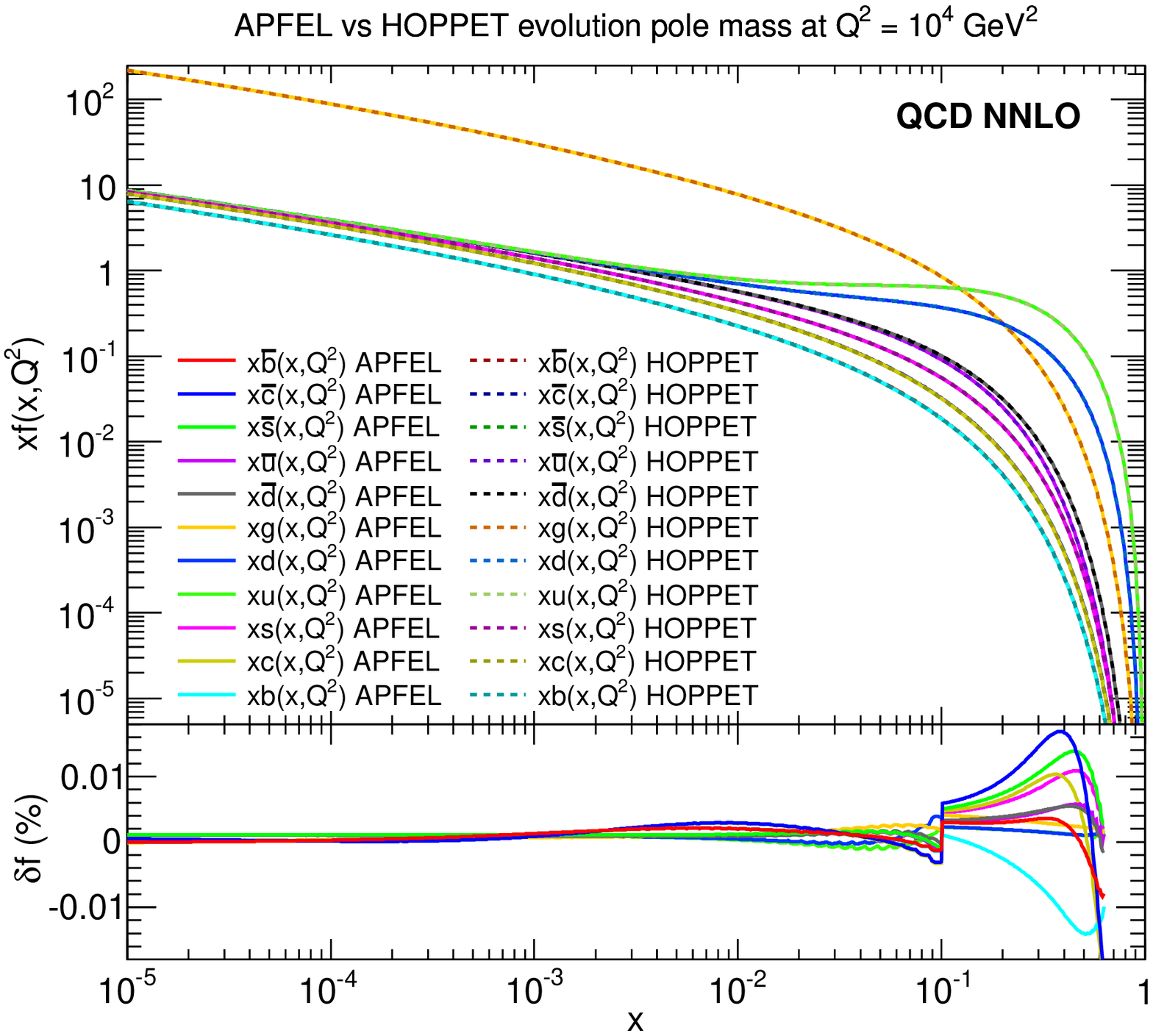}\includegraphics[scale=0.4]{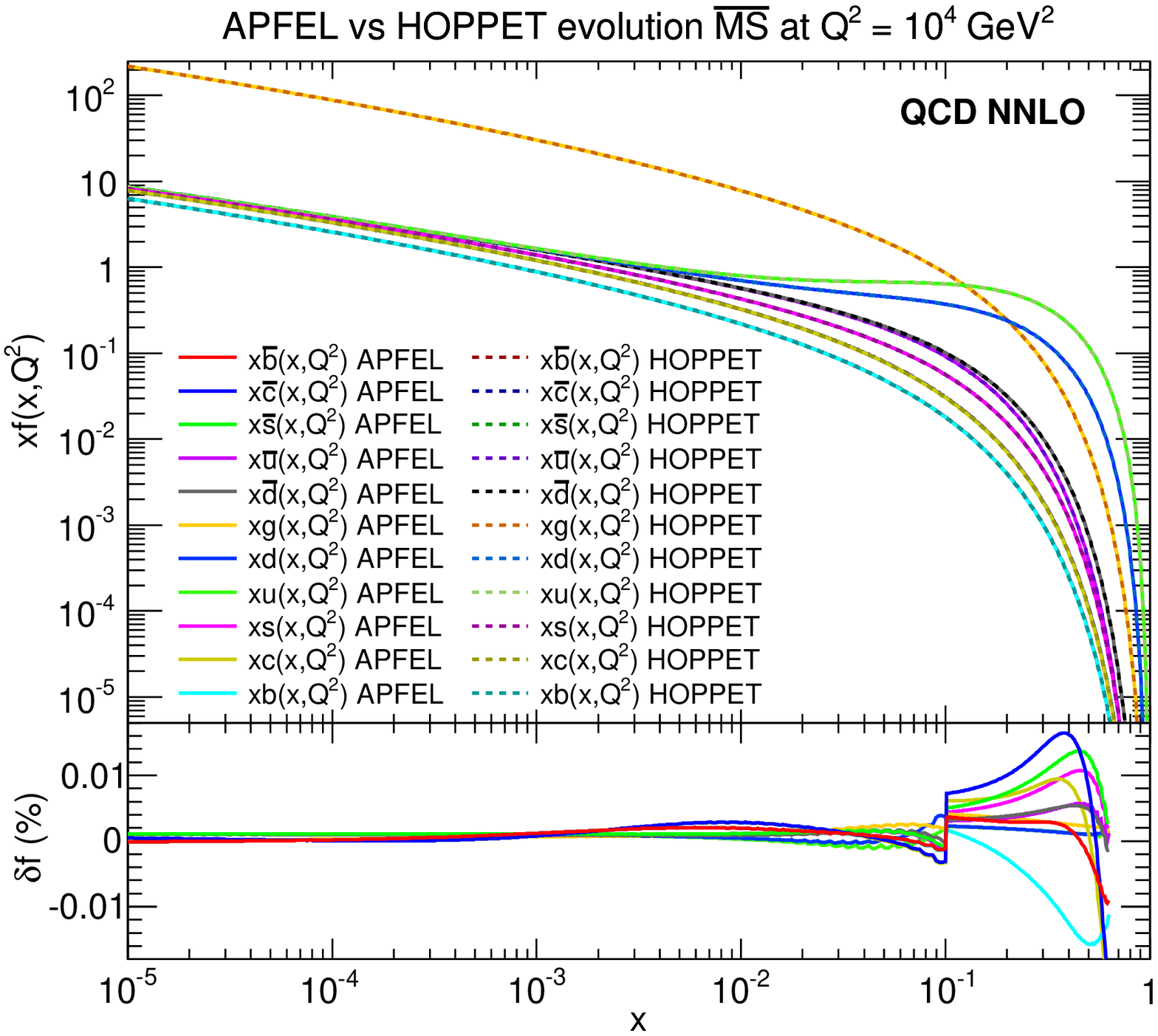}
\caption{\small Comparison between PDFs evolved at NNLO in QCD
 using {\tt APFEL}
  and {\tt HOPPET}, from $Q_0^2=$2 GeV$^2$ up to
  $Q^2=$10$^4$ GeV$^2$, using the Les Houches  PDF benchmark settings.
The comparison is performed 
 in the pole mass scheme (left) and  in the $\MSbar$ scheme (right).
The lower plots show the percent differences between the two
codes.
}
\label{fig:APFELvsHOPPET-one}
\end{figure}

To begin with, we validate the QCD evolution in {\tt APFEL} by
comparing it
with the results from the {\tt HOPPET} program,
 version 1.1.5, up to NNLO, and
 using both  pole and $\MSbar$ heavy quark masses.
The settings are the same as in the original Les Houches PDF
evolution benchmark~\cite{Dittmar:2005ed}.
In the case of  $\MSbar$ masses, we take the $\MSbar$ 
Renormalization-Group-Invariant charm mass
$m_c(m_c)$ to have the same
numerical values as the pole masses.
In all the comparisons in this section,
the interpolation settings in {\tt APFEL} are the default ones
discussed in Sect.~\ref{sec:methods}.

Results for the evolved PDFs at $Q^2=10^{4}$ GeV$^2$ for both
{\tt HOPPET} and {\tt APFEL} are shown in Fig.~\ref{fig:APFELvsHOPPET-one}.
The left plot shows the results using pole masses, while the right
plot corresponds to the case of $\MSbar$ masses.\footnote{In this
  latter case, the predictions from {\tt HOPPET} were also compared with those
of the internal NNPDF code {\tt FastKernel} in~\cite{Bertone:2012wi}
finding good agreement.} Fig.~\ref{fig:APFELvsHOPPET-one} also
shows the percent difference between both predictions, to show the
excellent agreement obtained for the whole range in $x$, being at most
$\sim 0.02\%$ at large-$x$, where PDFs have more structure.

\subsection{QED evolution}

We present now the numerical comparison of {\tt APFEL} with three 
different QCD$\otimes$QED parton evolution codes: first
{\tt partonevolution}, then the internal evolution program
used in the MRST04QED analysis, and finally the
{\tt FastKernel} program used in the NNPDF2.3QED analysis

\subsubsection{Comparison with {\tt partonevolution}}

To begin with, we compare the results for the 
coupled QCD$\otimes$QED DGLAP evolution in {\tt APFEL} with 
those of the public  {\tt partonevolution} code~\cite{Weinzierl:2002mv,Roth:2004ti}, version 1.1.3.
%

\begin{figure}[t]
\centering
\includegraphics[scale=0.4]{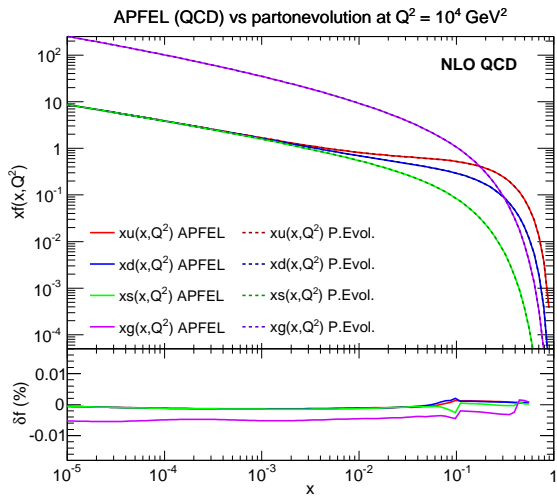}
\caption{\small Comparison between PDFs evolved at NLO in QCD (without QED
corrections) with
  {\tt APFEL} and {\tt partonevolution}, from $Q_0^2=4$ GeV$^2$ up
to $Q^2=10^4$ GeV$^2$.
 The same settings of the PDF benchmark study of
Ref.~\cite{Blumlein:1996rp} have been used.
The lower plot shows the percent differences between the two codes.
}
\label{fig:APFELvsPARTONEVOLUTIONQCD}
\end{figure}

\begin{figure}[h]
\centering
\includegraphics[scale=0.36]{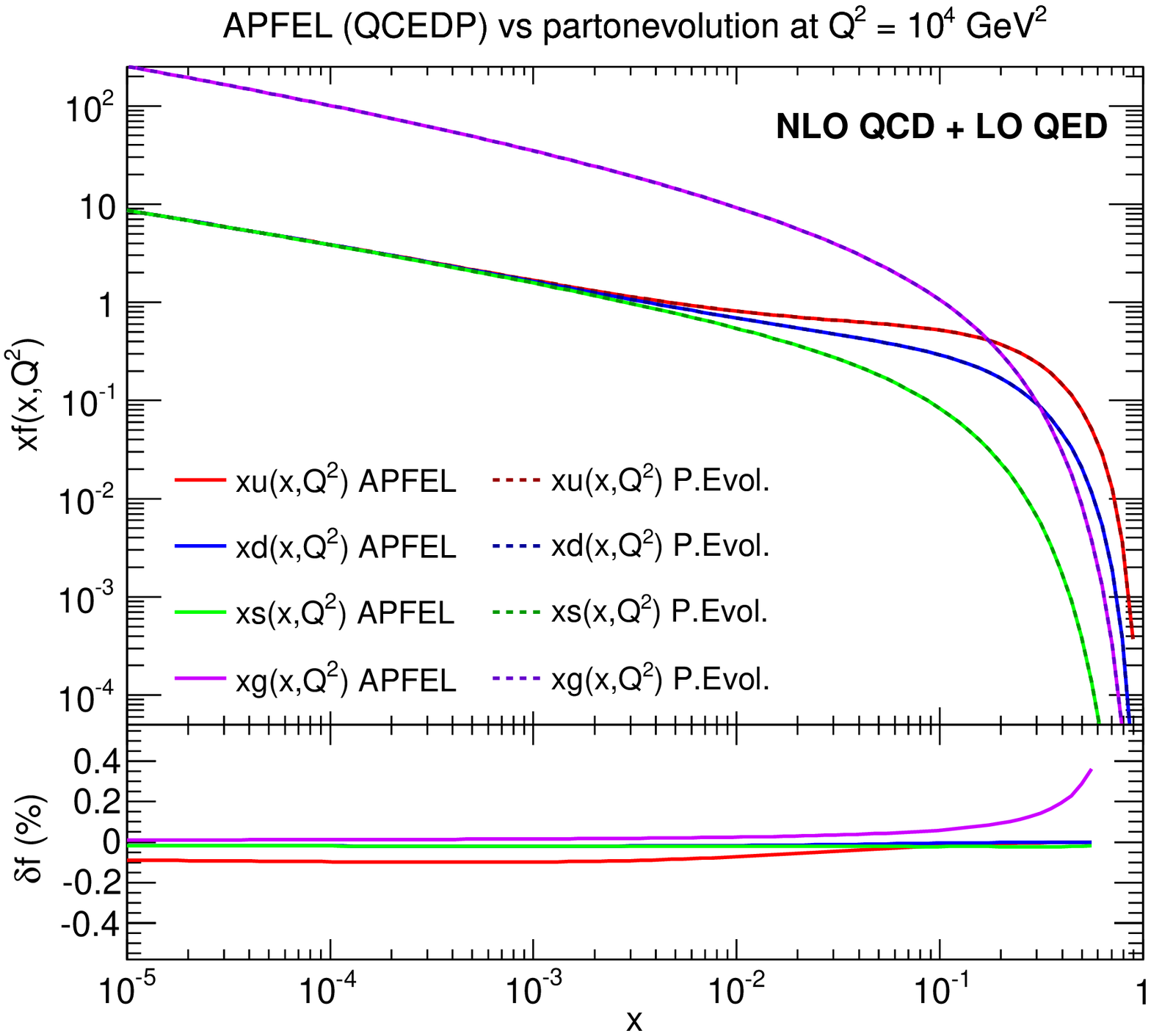}\includegraphics[scale=0.36]{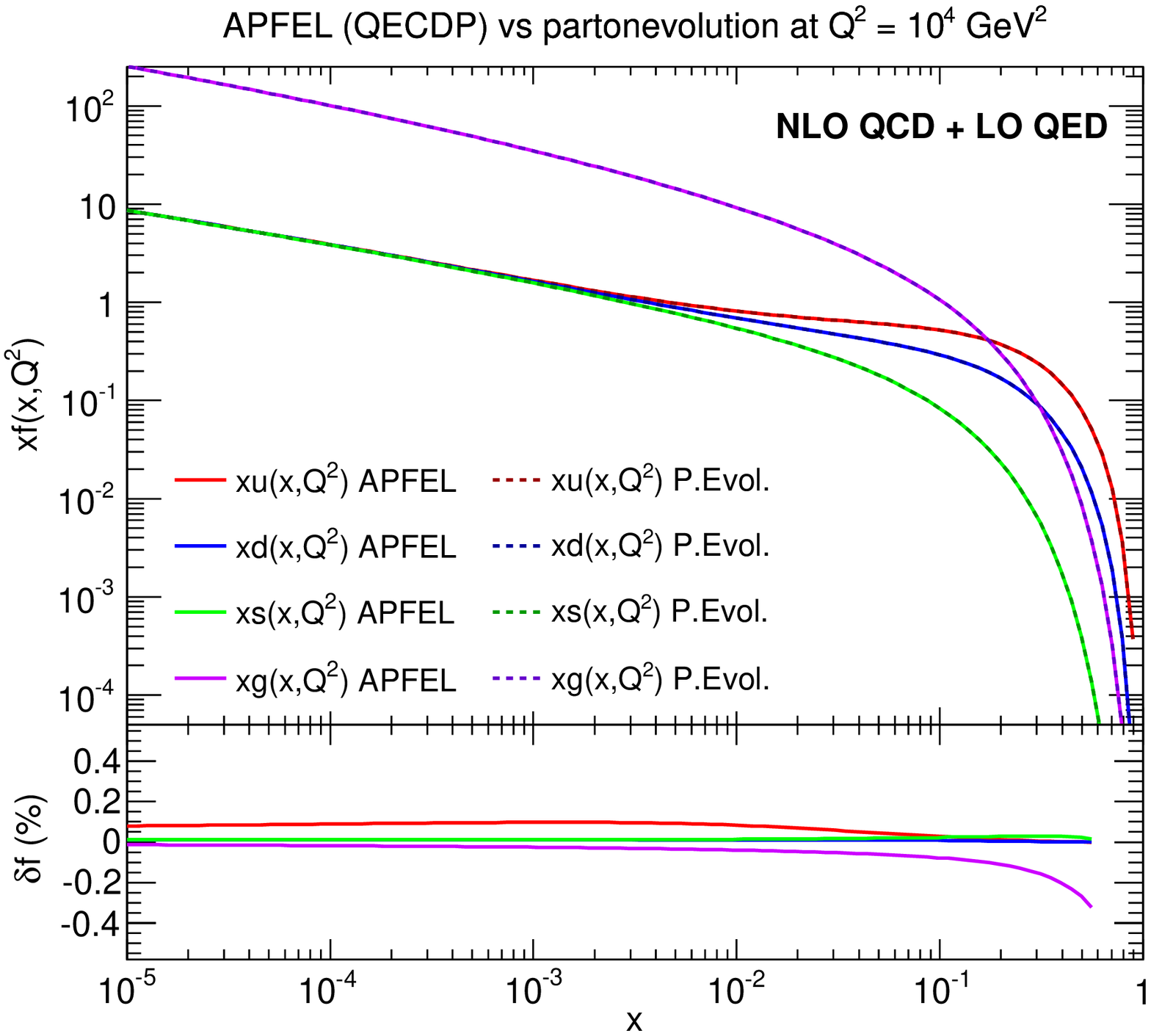}
\includegraphics[scale=0.36]{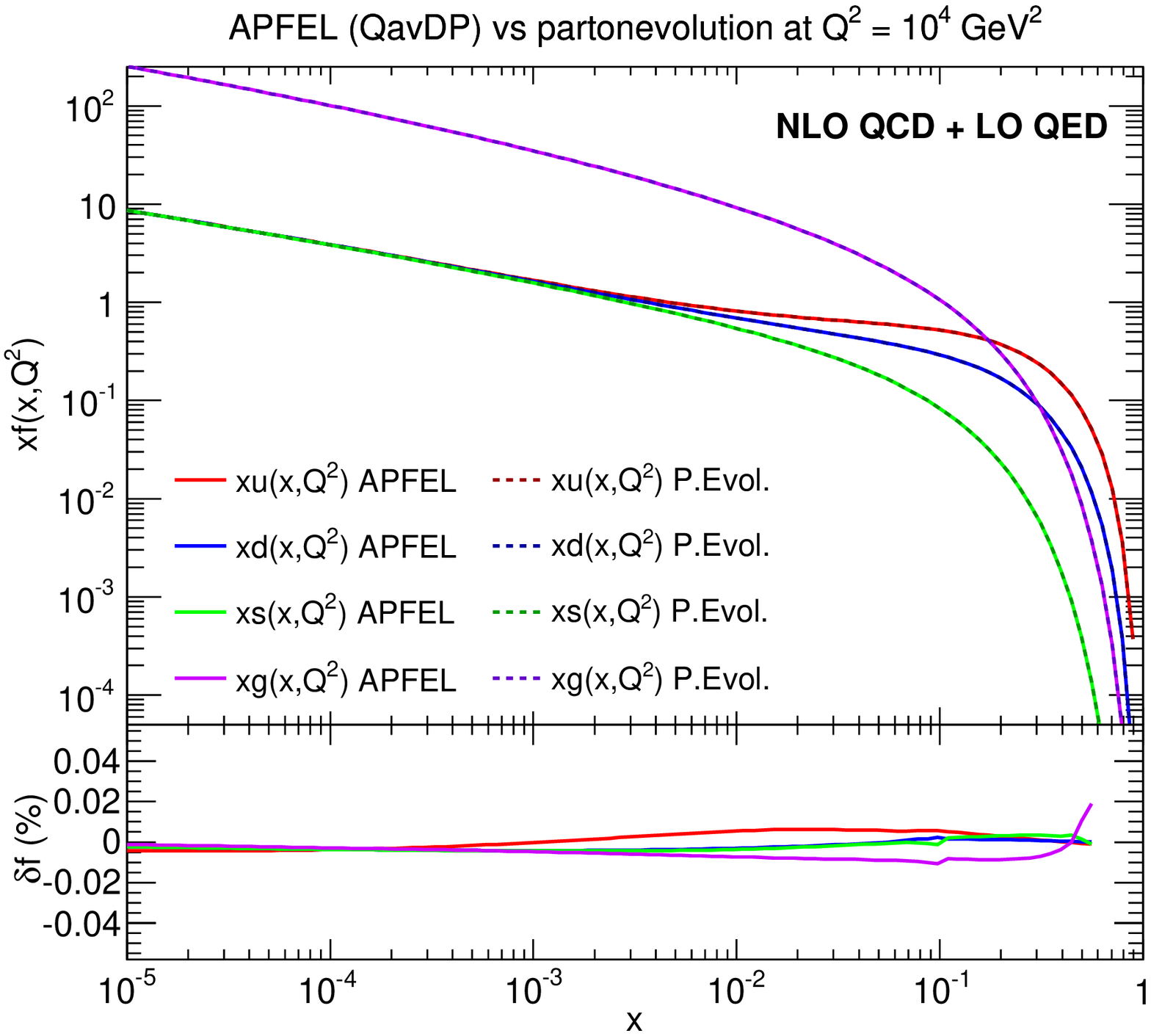}\includegraphics[scale=0.36]{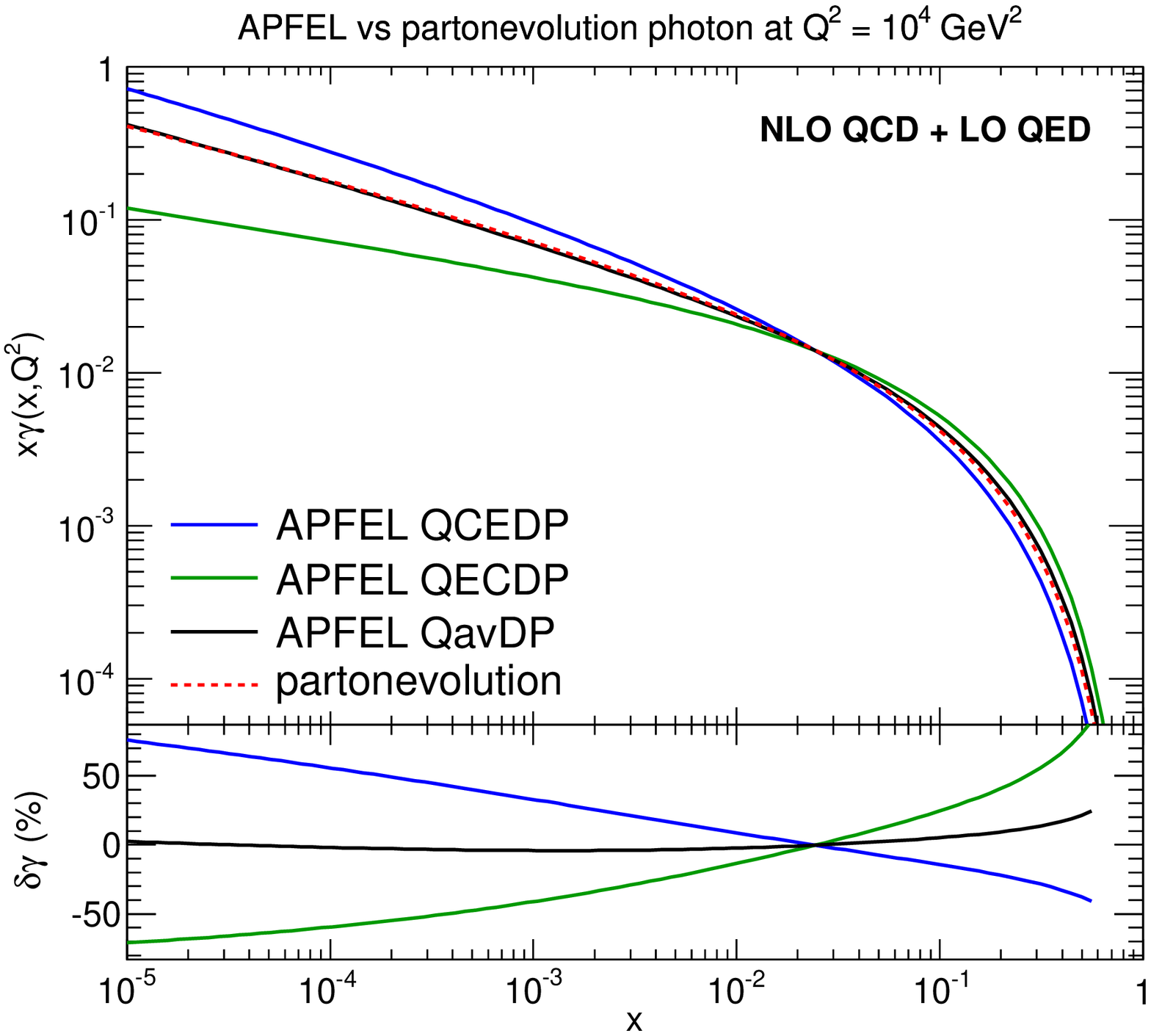}
\caption{\small Comparison between PDFs evolved at NLO in QCD
and LO and QED using {\tt APFEL}
  and {\tt partonevolution}, from $Q_0^2=$4 GeV$^2$ up 
$Q_0^2=$10$^4$ GeV$^2$
The same settings of the PDF benchmark study of
Ref.~\cite{Blumlein:1996rp} have been used.
We show the comparison of quark and gluon PDFs using in
{\tt APFEL} the {\tt QCEDP} solution (upper left plot), the {\tt QECDP}
solution (upper right plot), the {\tt QavDP} solution (lower
left plot) and then the photon PDF $\gamma(x,Q^2)$
in the two codes for the different {\tt APFEL} options.
For each comparison, we also show the percent differences with respect
to the {\tt partonevolution} results.
}
\label{fig:APFELvsPARTONEVOLUTION}
\end{figure}

To perform the benchmark, we use {\tt APFEL}
with the same settings used in the original
publication~\cite{Roth:2004ti} to present the numerical results of
{\tt partonevolution}, \textit{i.e.}
we take the input PDFs from the  toy model used in the benchmarking exercise of Ref.~\cite{Blumlein:1996rp},
given by:
\bea
xu_v(x)=A_ux^{0.5}(1-x)^3\, , \quad &\,& xd_v(x)=A_dx^{0.5}(1-x)^4 \, ,\nonumber \\
xS(x)=A_Sx^{-0.2}(1-x)^7 \, ,\quad &\,& xg(x)=A_gx^{-0.2}(1-x)^5 \, , \nonumber \\
xc(x)=0 \, , \quad &\,& x\bar{c}(x)=0 \, , 
\eea
at the initial scale $Q_0^2=4$~GeV$^2$, with a SU(3) symmetric sea that carries 15\% of the proton's momentum
at $Q_0^2$, and only four active quarks are considered even above the bottom threshold.
This toy model should not be confused with that used in the Les Houches
PDF benchmark study, used elsewhere in this paper.
In addition, 
the photon PDF is set to zero at the initial scale, that is $\gamma(x,Q_0^2)=0$.

In order to set up the baseline, we ran the two codes at NLO QCD only, 
switching off the QED corrections. As can be seen from
Fig.~\ref{fig:APFELvsPARTONEVOLUTIONQCD}, good agreement is achieved.
We can then move to the combined QCD$\otimes$QED evolution. Results are summarized in Fig.~\ref{fig:APFELvsPARTONEVOLUTION}, where
we compare the evolution of quark, gluon and
photon PDFs given by the two codes, using the three different options for
the solution of the coupled equations provided by {\tt APFEL}: {\tt QCED}, {\tt QECD}
and {\tt QavD} (see Sect.~\ref{sec-theory}).
Note that in the fixed-flavor-number scheme the distinction
between ``series'' and ``parallel'' solutions is immaterial.
As expected, the best agreement between the two codes is obtained with the
{\tt QavD} option. As discussed in Sect.~\ref{sec-theory},
this option ensures that the methods used for the solution of the evolution
equations in the two codes differ only by $\mathcal{O}\lp \alpha^2\rp$ 
terms rather than by the larger $\mathcal{O}\lp \alpha \alpha_s\rp$
corrections. 
With these settings the evolution of quarks and gluon is essentially
identical, with differences at most being $\mathcal{O}\lp 0.01\%\rp$,
while differences in the evolution of $\gamma(x,Q^2)$ are below
the few percent level except at the largest values of $x$.

The other two options, {\tt QCED} and {\tt QECD}, lead to differences in the PDF
evolution of order $\mathcal{O}\lp \alpha \alpha_s\rp$
as compared to the {\tt partonevolution} solution, corrections
that, while being formally subleading, can be numerically large.
For quark and gluon PDFs, the largest differences are seen for the gluon
PDF, which can be up to 0.5\% at the largest values of $x$.
More substantial differences appear for the photon PDF, where of course
higher-order QED effects are expected to be sizable.
In the case of $\gamma(x,Q^2)$, the three formally equivalent solutions
can differ by up to 70\%, both at small and large-$x$.
As will be discussed below in Sect.~\ref{sec:consistency}, these large
differences of the {\tt QCED} and {\tt QECD} solutions arise from unresummed large
logarithms of the QCD and QED factorization scales. However, let us also recall that the order of magnitude is similar to that obtained
in the comparison of LO and NLO evolution for the gluon PDF in QCD~\cite{Ball:2011uy}.

\subsubsection{Comparison with MRST04QED}

Another instructive comparison is provided by the QED evolution used in the
determination of the MRST04QED parton distributions~\cite{Martin:2004dh}.
Though the original evolution code is not publicly available, the
evolution which was used can be indirectly accessed via the public
{\tt LHAPDF} grids.
In this case, it is not possible to use the Les Houches
benchmark settings, and we are instead forced to use the same boundary
conditions for the PDFs at $Q_0$ as those used in the MRST04QED fit
as well as the same values of the heavy quark masses and reference
coupling constants.
The available MRST04QED fit was obtained at NLO in QCD in the VFN scheme,
therefore it is possible to perform a meaningful
comparison with the results of their
evolution by using {\tt APFEL} at NLO with the same
settings.

\begin{figure}[h]
\centering
\includegraphics[scale=0.36]{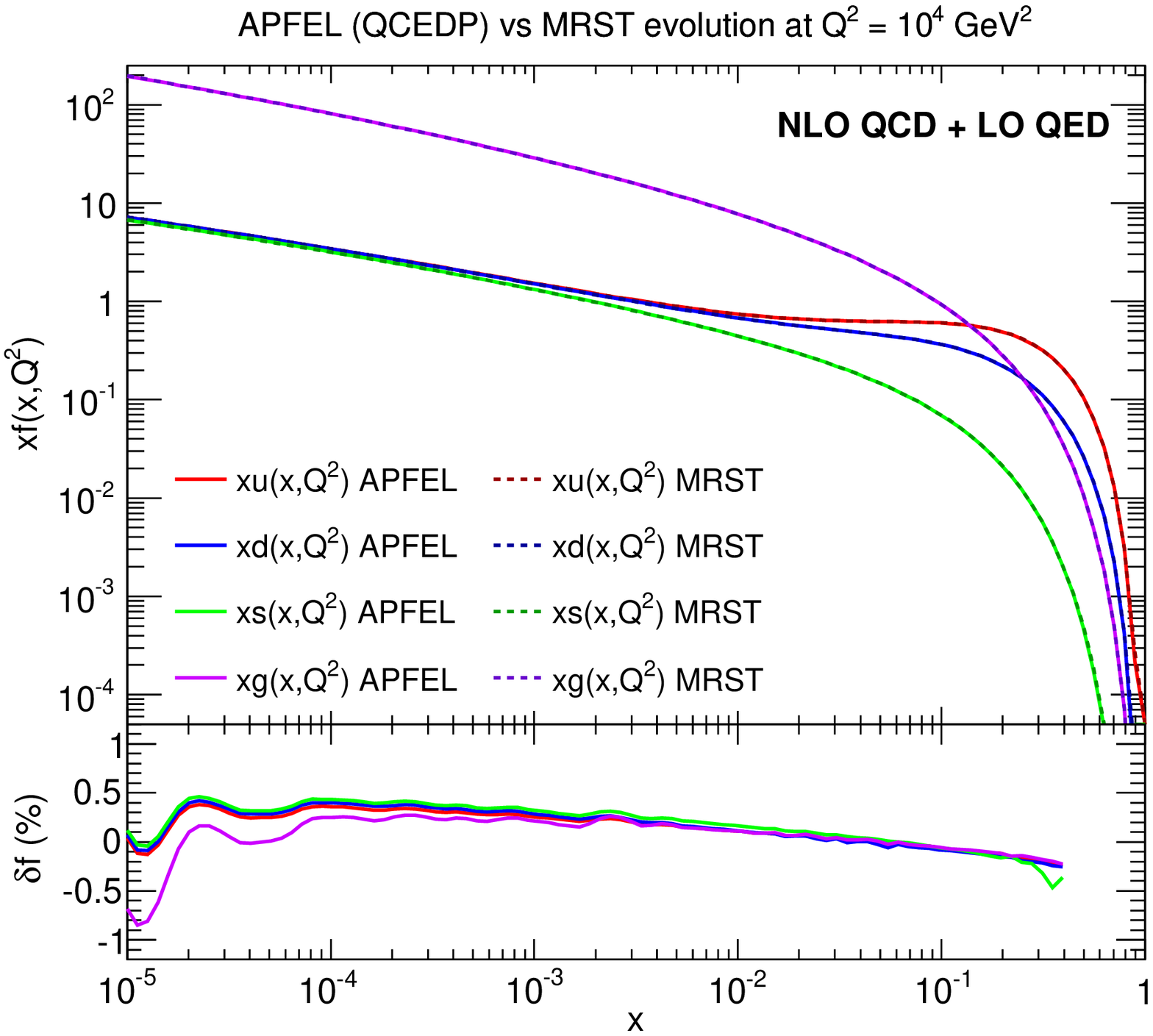}\includegraphics[scale=0.36]{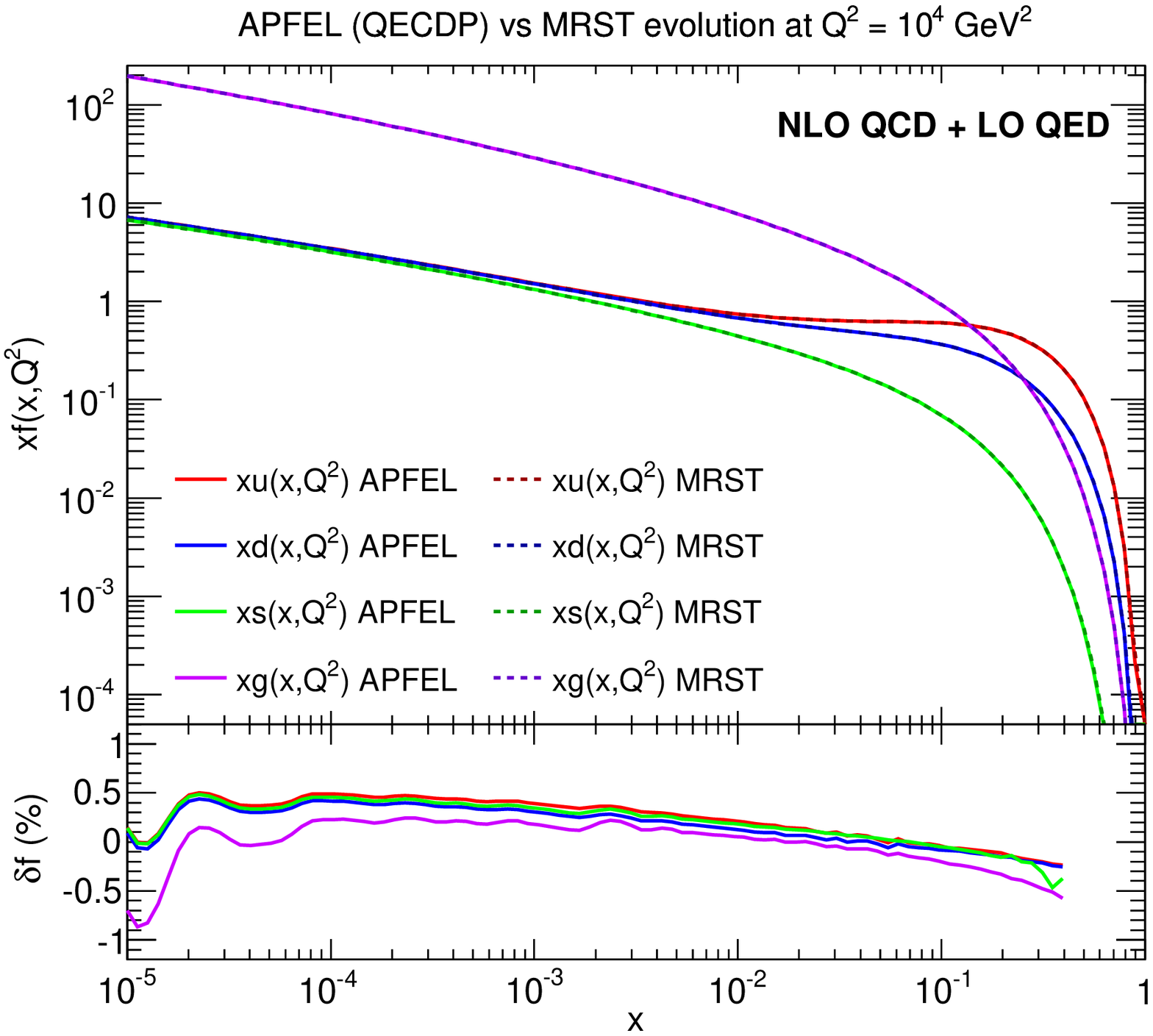}
\includegraphics[scale=0.36]{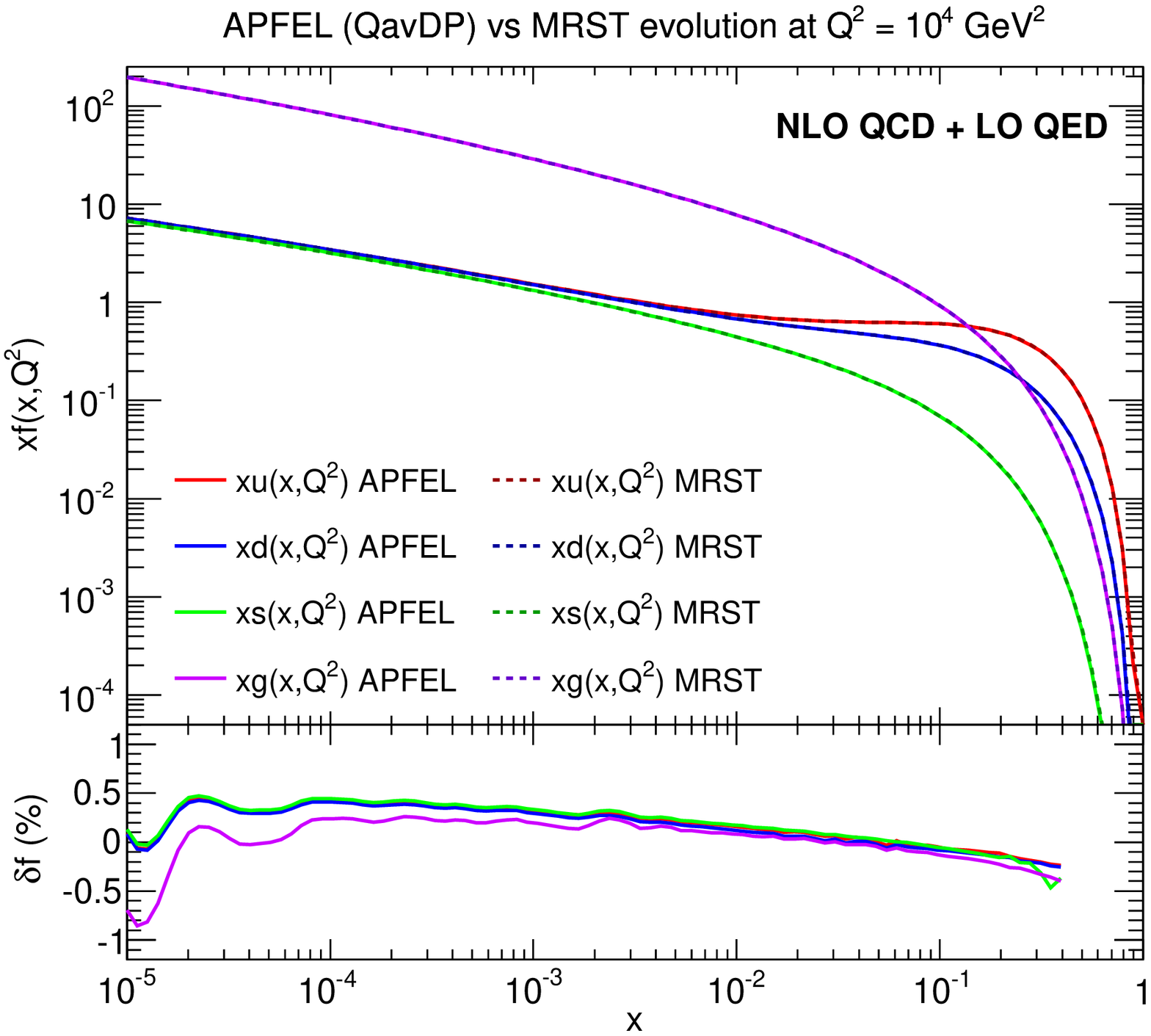}\includegraphics[scale=0.36]{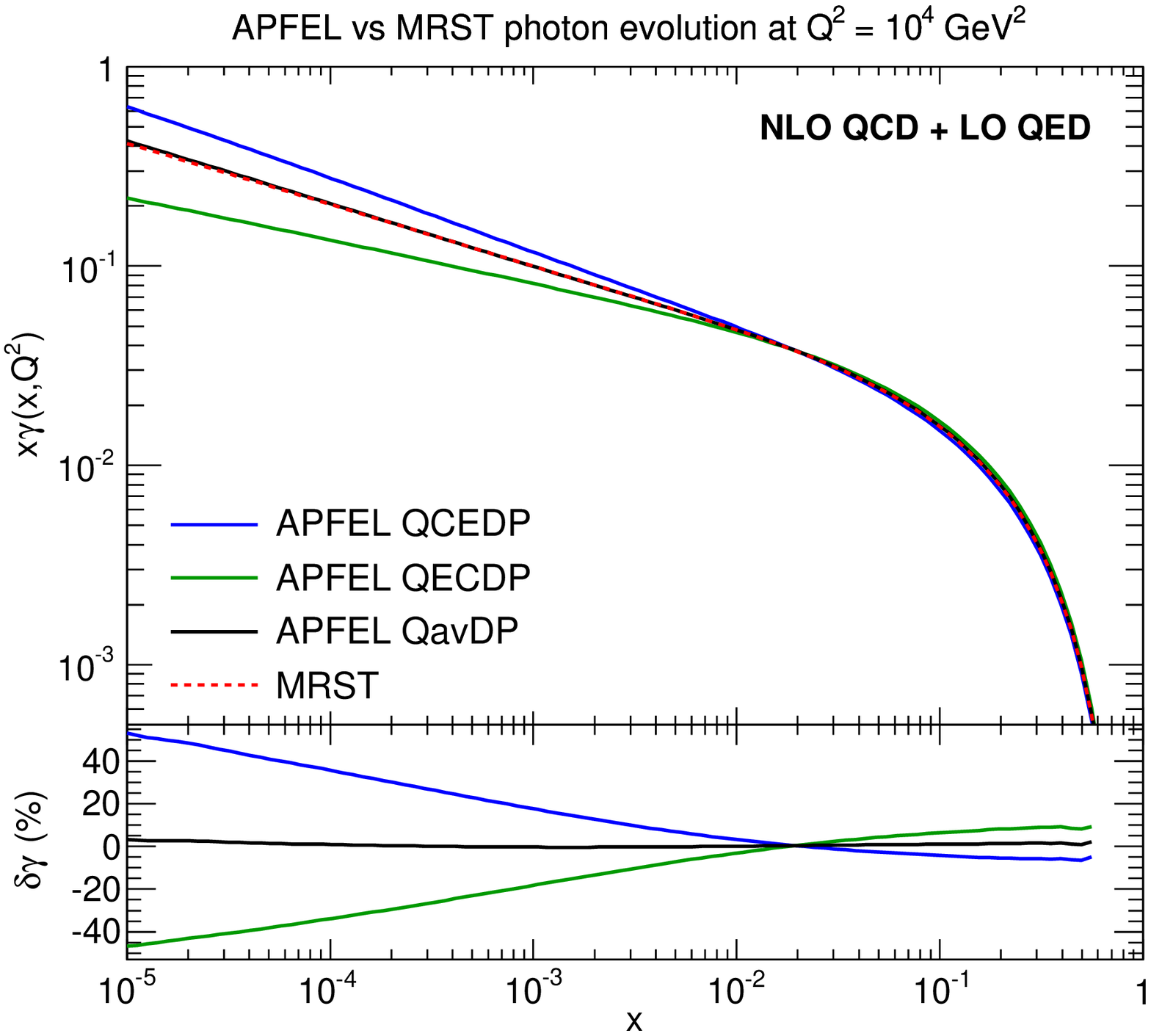}
\caption{\small Comparison between PDFs evolved using {\tt APFEL} and the
  internal MRST04QED parton evolution, from $Q_0^2=$2 GeV$^2$ up to $Q^2=$10$^4$
  GeV$^2$.
The boundary conditions for the PDFs are the same as
those of the 
MRST04QED fit.
PDF evolution is performed at NLO in QCD and LO in QED, in the 
variable-flavor number
scheme.
We show the comparison of quark and gluon PDFs using in
{\tt APFEL} the {\tt QCEDP} solution (upper left plot), the {\tt QECDP}
solution (upper right plot), the {\tt QavDP} solution (lower
left plot) and then the photon PDF $\gamma(x,Q^2)$
in the two codes for the different {\tt APFEL} options.
}
\label{fig:APFELvsMRST-one}
\end{figure}

The comparison between the {\tt APFEL} predictions and the MRST04QED
evolution is shown in  Fig.~\ref{fig:APFELvsMRST-one}.
PDFs have been evolved using {\tt APFEL} and the internal MRST
evolution from $Q_0^2=$2 GeV$^2$ up to $Q^2=$10$^4$ GeV$^2$.
We have explored the same three different options for the
QCD$\otimes$QED combined evolution provided by {\tt APFEL} as in the
comparison with {\tt partonevolution}, which differ only in the treatment of
formally subleading terms.
As can be seen from Fig.~\ref{fig:APFELvsMRST-one}, for the
evolution of the photon PDF the best agreement between {\tt APFEL} and
MRST04QED is achieved when the {\tt QavDP} solution is used.
The differences in this case are 1\% at most
for $\gamma(x,Q^2)$ and much smaller for quark and gluon PDFs.
It is also clear from  Fig.~\ref{fig:APFELvsMRST-one} that
other different options for solving the evolution equations
in the presence of QED effects have a small impact
on the evolution of quark and gluon PDFs, but for the photon
PDF they can lead to differences up to 40\% at small-$x$,
for the reasons explained above.
On the other hand, it should be taken into account that theoretical uncertainties in the QED evolution
are subdominant with respect to the experimental
uncertainties in the photon PDF at small-$x$~\cite{Ball:2013hta},
due to the lack of experimental constraints.

\subsubsection{Comparison with {\tt FastKernel}}

The third validation test of the QED evolution
 is provided by comparing the results of {\tt APFEL}
with the predictions given by the {\tt FastKernel} internal
NNPDF code~\cite{Ball:2010de}.
This comparison is illustrative since the {\tt FastKernel} code was used
in the derivation of the NNPDF2.3QED sets, presented in
Ref.~\cite{Ball:2013hta} and where the {\tt FastKernel} was compared to
the {\tt partonevolution} code finding reasonable agreement, with
some small differences arising from a different treatment of the subleading
terms.
Now we revisit this issue using the {\tt APFEL} flexibility to explore
different options for the solution of the QCD$\otimes$QED coupled evolution equations.

\begin{figure}[h]
  \begin{centering}
    \includegraphics[scale=0.5]{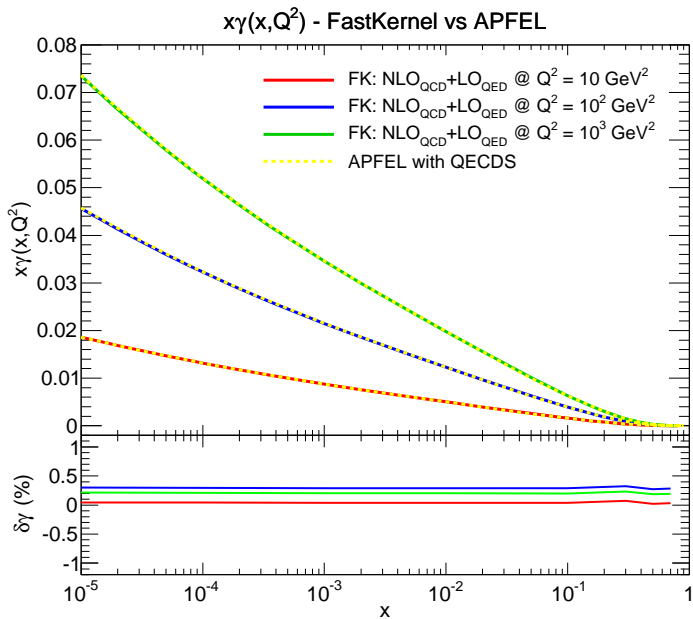}
    \par\end{centering}
  \caption{\small \label{fig:APFELvsFASTKERNEL} The evolution of the
photon PDF $\gamma(x,Q^2)$ at NLO in QCD and LO in QED for
various values of $Q^2$, performed both with {\tt APFEL} and
with {\tt FastKernel}.
The Les Houches benchmark settings have been used, 
supplemented by the boundary condition $x\gamma(x,Q^2_0)=0$, with $Q^2_0=2$~GeV$^2$.
For {\tt APFEL}, the {\tt QECDS} solution has been used.
The lower plot shows the percent difference between the two calculations.
}
\end{figure}

%
The basic strategy underlying the {\tt FastKernel} code~\cite{Ball:2008by} is to solve
the evolution equations in Mellin space and then invert the evolution kernels
back to $x$ space, where they are convoluted with
the initial $x$-space PDFs to obtain the evolved PDFs.
The NNPDF2.3QED sets use the truncated solution for the Mellin
space DGLAP equations, and so cannot directly be compared with
{\tt APFEL}, which is based instead on the expanded solution
(as any $x$-space code). 
Therefore, for the purpose of this benchark comparison, we
have produced the combined NLO QCD and LO QED  
predictions with the {\tt FastKernel} code using the
 expanded solution of the Mellin space
equations, and compared them to the
{\tt APFEL} results.

The comparison is presented in Fig.~\ref{fig:APFELvsFASTKERNEL}.
We show the evolution of the
photon PDF $\gamma(x,Q^2)$ at NLO in QCD and LO in QED for
various values of $Q^2$, performed both with {\tt APFEL} and
with the {\tt FastKernel} code.
For {\tt APFEL}, the {\tt QECDS} solution has been used: this is
indeed equivalent
 to the procedure that has been used in the NNPDF2.3QED fits.
The Les Houches benchmark settings have been adopted, 
supplemented by the boundary condition $\gamma(x,Q^2_0)=0$, with $Q^2_0=$2 GeV$^2$.
In the lower plot of  Fig.~\ref{fig:APFELvsFASTKERNEL} we 
provide the percent differences between the two calculations.
As can be seen, differences are small, at the few per-mil level, 
showing that the two codes are in good agreement when common settings are adopted.

In summary, the QCD$\otimes$QED evolution used in the NNPDF2.3QED fits differs
from that used in other codes, such as {\tt partonevolution} and MRST04QED, only
by higher-order terms: truncated versus iterated solution of the evolution
equations in Mellin space and different treatment of the subleading $\mathcal{O}\lp \alpha \alpha_s\rp$ terms.
As has been shown in Figs.~\ref{fig:APFELvsPARTONEVOLUTION} and~\ref{fig:APFELvsMRST-one},
while for quark and gluon PDFs these differences are small, 
for the photon PDF these inherent theoretical uncertainties are 
substantial and could possibly be reduced including higher-order
corrections in the QED coupling in the combined
evolution equations.
Such improvements might be required by future, more precise experimental data.

\subsection{Consistency of the coupled solution}
\label{sec:consistency}

Now we turn to discuss the consistency of the procedure
adopted in {\tt APFEL} to combine the solutions
of the QCD and QED evolution equations.
As presented in Sect.~\ref{sec-theory}, the combination of QCD and QED
solutions in {\tt APFEL}  is perfomed
 at the level of the evolution operators, rather than
at the level of splitting functions, as done, for instance, in the {\tt
  partonevolution} and  MRST04QED codes. 
This procedure leads to the introduction of two
different factorization scales, $\mu_{\rm QCD}$ and $\mu_{\rm QED}$, 
which in principle are allowed to vary in a fully independent way.

A possible objection to this approach is that,
in the case in which
 $\mu_{\rm QCD}$ and $\mu_{\rm QED}$ are very different from each
other, this procedure might lead to the presence of numerically large,
unresummed logarithms.
On the other hand, in Sect.~\ref{sec-theory} we showed
that these terms are $\mathcal{O}\lp
\alpha \alpha_s\rp$ for the {\tt QECD} and the {\tt QCED} solutions and
$\mathcal{O}\lp
\alpha^2\rp$ for the {\tt QavD} solution, that are both perturbatively subleading.

In order quantify the impact of these potentially large (subleading)
logarithms, we have performed with {\tt APFEL} the combination of QCD and QED
evolutions not over the whole (possibly large) $\lc Q_0,Q\rc$ range, but
rather dividing it in small intervals
$\lc Q_0,Q_1\rc$, $\lc Q_1,Q_2\rc$, $\ldots$, 
$\lc Q_N,Q\rc$, and performing the combination on each interval. 
This procedure ensures that no artificially large logarithm of two
widely different scales appears in the solution.
An example of how to construct this small-step approximation with
the available functions of {\tt APFEL} has been included in the {\tt example} folder of the main source
folder.

As an illustration, using again the settings of the Les Houches PDF
evolution benchmark~\cite{Dittmar:2005ed}, 
supplemented by the ansatz  $\gamma(x,Q_0) = 0$,
 we have evolved the PDFs 
between $Q_0^2 = 2$ GeV$^2$ and
$Q^2 = 10000$ GeV$^2$.
We  have then compared the standard, single-step solution with the new
solution sketched above and based on combining
the results of many small evolution steps.
As expected, no significant differences were observed for quarks and
gluon, so in the following we concentrate on the different results for the 
evolution of the photon PDF.

\begin{figure}[h]
\centering
\includegraphics[scale=0.41,angle=270]{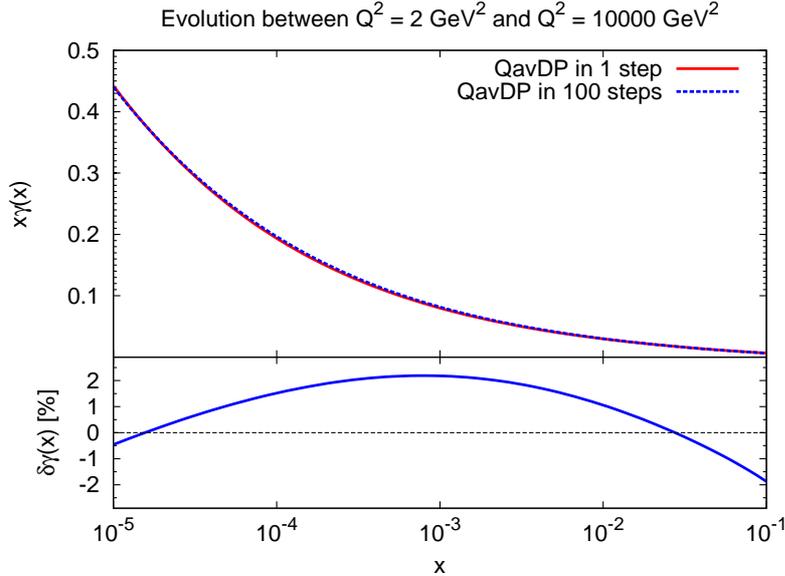}
\caption{\small Evolution of the
photon PDF $\gamma(x,Q^2)$ between $Q_0^2=2$ GeV$^2$ and
10$^4$ GeV$^2$ with the {\tt QavDP} solution.
We compare the evolution performed
with a single step with that obtained with 100
steps logarithmically spaced in $Q^2$.
The lower panel shows the percentage differences between the two
methods.
}
\label{fig:StepSolution3}
\end{figure}

In Fig.~\ref{fig:StepSolution3} we compare 
the standard {\tt QavD} solution, performed with a single
step between $Q_0^2$ and $Q^2$, with the corresponding
solution where the full range $\lc Q_0^2, Q^2\rc$ has been divided
into 100 logarithmically spaced intervals.
%
In the bottom  panel of Fig.~\ref{fig:StepSolution3}
we show the percentage difference between the two results:
reasonable agreement is found, with small residual differences
at the level of 2\% at most.
This comparison confirms that the {\tt QavDP} solution adopted in {\tt APFEL}
is free of numerically large scale logarithms and that a single
$Q^2$ interval leads to reliable results.
The small impact of potentially large logarithms can be explained
considering that they are suppressed by a factor $\alpha^2$.

\vspace{0.3cm}

\begin{figure}[h]
\centering
\includegraphics[scale=0.41,angle=270]{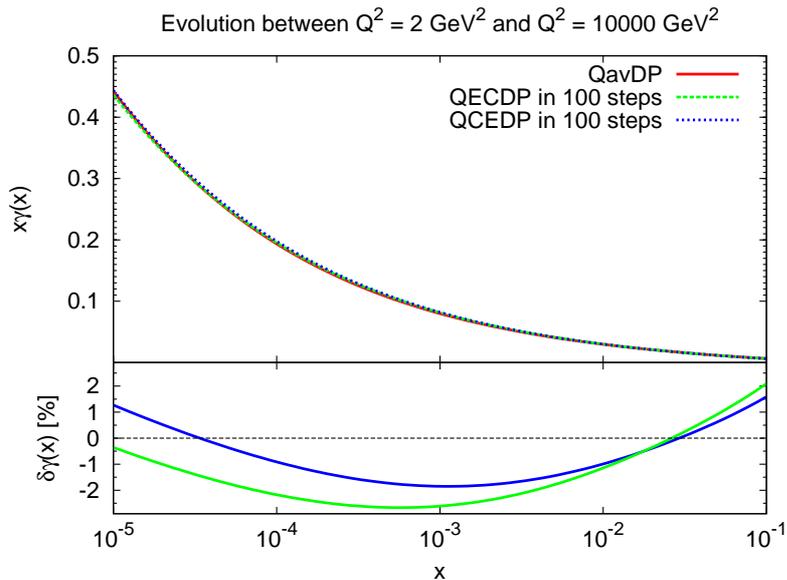}
\caption{\small 
Comparison of the evolution of the
photon PDF $\gamma(x,Q^2)$ between $Q_0^2=2$ GeV$^2$ and
10$^4$ GeV$^2$ obtained with the {\tt QavDP} solution with a single step, with those of the
{\tt QECDP} and {\tt QCEDP} solutions with 100 logarithmically spaced steps.
The lower panel shows the percentage differences between the two
methods.
}
\label{fig:StepSolution1}
\end{figure}

Once we have explicitly verified
 that no large unresummed logarithms are present in the {\tt QavD} solution, we turn to consider the {\tt QECD} and
the {\tt QCED} solutions.
The results for the 100-step evolution of the photon PDFs in these two cases are shown in 
Fig.~\ref{fig:StepSolution1} and compared to the one-step {\tt QavD} solution.
We see that, unlike the case where a single step is used
(see Fig.~\ref{fig:APFELvsMRST-one}) and where differences up to
40\% were observed (though only for small values of $x$), 
now differences between the three solutions
are 2\% at most.
This suggests that the $\mathcal{O}\lp \alpha\alpha_s\rp$ logarithmic terms that
affect the {\tt QECD} and the {\tt QCED} solutions are numerically
large and may spoil the respective perturbative evolution unreliable
if the evolution interval is too wide.

Therefore, we can conclude that the {\tt QavD} solution of the
coupled QCD$\otimes$QED evolution equations is the
most reliable one, since in this case potentially large unresummed
scale logarithms are absent.
On the other hand, this analysis also shows that the
{\tt QECD} and the {\tt QCED} solutions may introduce artificially large logarithms and
that an effective way to cancel them is to perform the
evolution in small steps combining sequentially the
results.
In this case, the results from the {\tt QECD} and the
{\tt QCED} solutions coincide
to a good approximation with that of the {\tt QavD} solution,
so that all three strategies lead to the same numerical solution.

\subsection{Deep-Inelastic Scattering structure functions}

\begin{figure}[h]
\centering
\includegraphics[scale=0.35,angle=270]{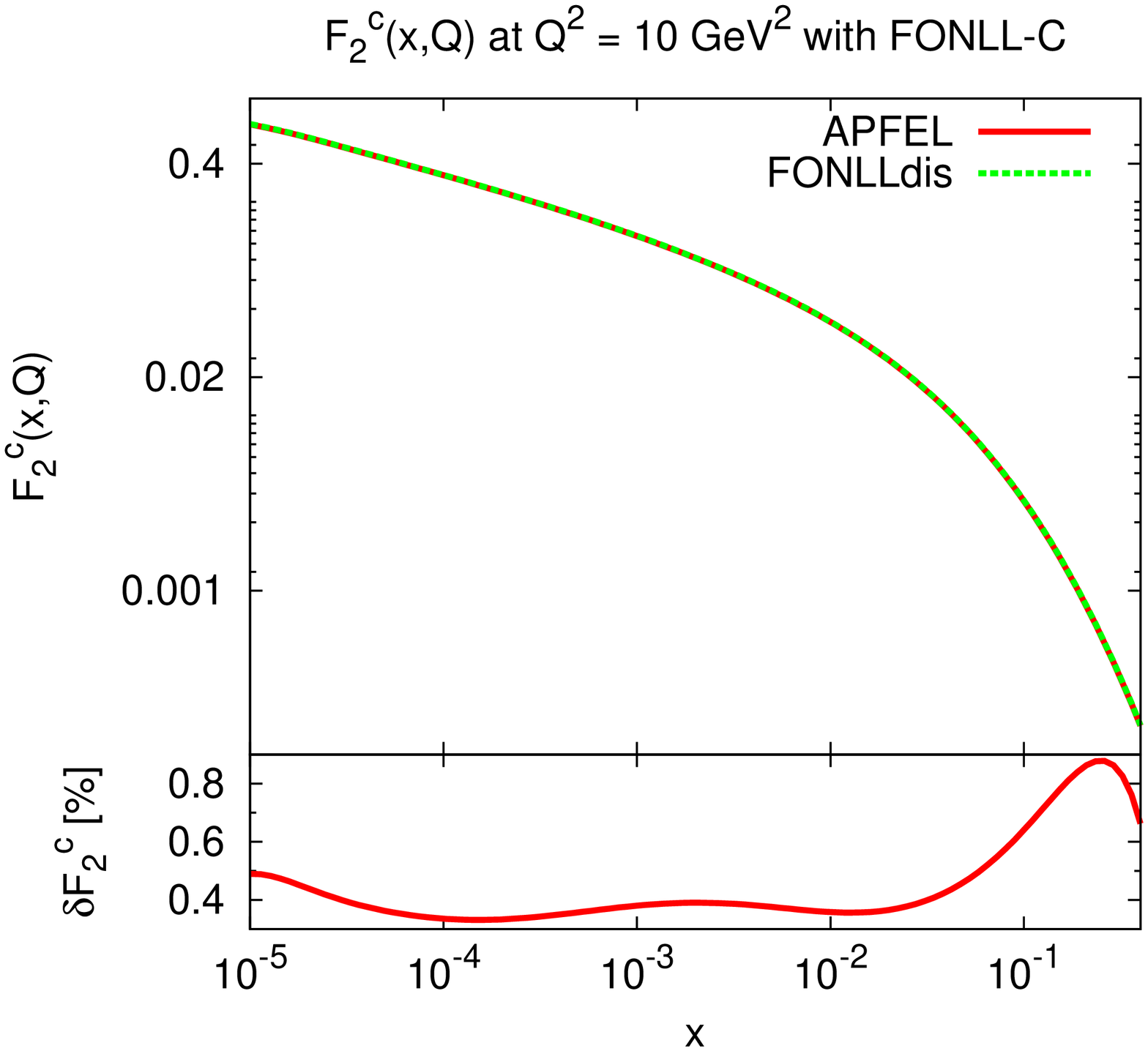}
\includegraphics[scale=0.35,angle=270]{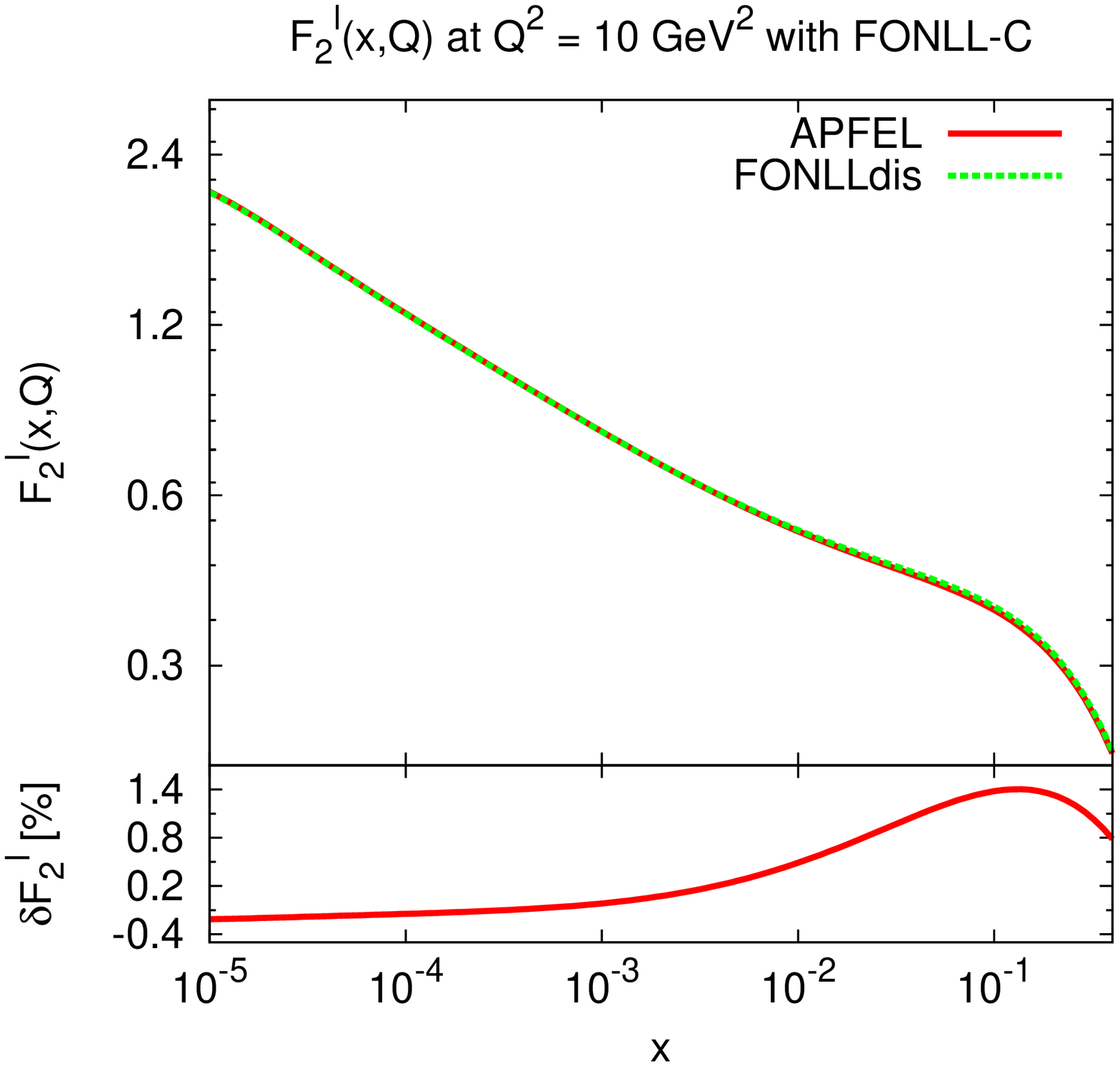}
\caption{\small Comparison between {\tt APFEL} and {\tt
    FONLLdis}~\cite{Forte:2010ta} for the charm
structure function $F_2^c$ (left) and the light structure function
$F_2^l$ (right) in the FONLL-C scheme in the range $x \in [10^{-5}:10^{-1}]$
at $Q^2 = 10$ GeV$^2$. The lower panel displays the
percentage different in percent between the two computations. \label{fig:APFELvsFONLLdis} }
\end{figure}

Finally, in this section we briefly document
 the benchmarking of the DIS module implemented
in {\tt APFEL} against the publicly
 available code {\tt FONLLdis}~\cite{Forte:2010ta}, which
computes the charm structure functions in the FONLL
GM-VFN scheme using the exact $x$-space $\mathcal{O}\lp \alpha_s^2\rp$
heavy quark coefficient functions. 
The {\tt FONLLdis} code provides predictions for the electromagnetic
structure functions $F_2^l$, $F_2^c$, $F_L^l$ and $F_L^c$ up to
$\mathcal{O}\lp \alpha_s^2\rp$ in the
FONLL scheme. In Fig.~\ref{fig:APFELvsFONLLdis} we compare the
predictions for $F_2^l$ (left plot) and $F_2^c$ (right plot)
at $Q^2 = 10$ GeV$^2$ obtained with {\tt APFEL} with those obtained with {\tt FONLLdis} using
the Les Houches heavy quark Benchmark settings~\cite{Binoth:2010ra}.
As can be seen from the bottom panel of Fig.~\ref{fig:APFELvsFONLLdis},
the relative differences between {\tt APFEL} and {\tt FONLLdis} are always well below the percent level for
$F_2^c$ and at most of 1.3\% for $F_2^l$: a more than reasonable accuracy for phenomenology.

\section{Conclusions and outlook}

\label{sec-conclusion}

{\tt APFEL} is a new PDF evolution library
 that combines NNLO QCD
corrections with LO QED effects in the solution of the DGLAP 
equations.
It is the first public evolution 
code that allows to perform the coupled QCD$\otimes$QED evolution
up to NNLO in $\alpha_s$ and LO in $\alpha$, both in the FFN and VFN
schemes, and using either pole or $\MSbar$ heavy quark masses.
It is fast, accurate, and flexible, and can be easily accessed through its
{\sc Fortran 77}, {\tt C/C++} and {\tt Python} interfaces.
From version {\tt  2.0.0}, the
{\tt APFEL} capabilities have been extended to include 
a new module that computes neutral- and
charged-current DIS observables
up to $\mathcal{O}(\alpha_s^2)$ in the FONLL scheme~\cite{Forte:2010ta}. In addition, we
provide a flexible user-friendly Graphical User Interface 
which provides access to all of the {\tt APFEL} functionalities without
the need of writing code and to produce high-quality plots in various
different formats.

Given the relevance of QED and electroweak corrections for precision
theoretical predictions at the LHC in the coming years, and
the corresponding needs for parton distributions that consistently include QED effects,
we believe that a library like {\tt APFEL} can become a useful tool for the PDF
fitting community.

In the short term, we plan to extend the capabilities of {\tt APFEL} to include
the solution of DGLAP equations with (N)NLO
 time-like splitting functions~\cite{Mitov:2006ic},
such as those required in global fits of fragmentation 
functions~\cite{deFlorian:2007hc}, as well as the evolution for 
polarized PDFs up to NLO, which could
be of interest in the determinations of spin-dependent 
parton distributions~\cite{Ball:2013lla,deFlorian:2009vb,Leader:2010rb}.
Another feature foreseen in future releases is the implementation of
different factorization schemes for the PDF
evolution, such as the DIS or the AB scheme, the latter useful in
the context of polarized fits~\cite{Altarelli:1998nb}.
In addition, while the current release of {\tt APFEL} allows only the
evolution of the proton PDFs, future releases will
also allow to simultaneously evolve both
proton and neutron PDFs.
This feature is important to enable dedicated studies of 
isospin symmetry violation,
both at low scales (non-perturbative) and at high scales, with an
additional component generated dynamically by the presence of QED effects
in the evolution~\cite{Ball:2013hta,Martin:2004dh}.

In a longer timescale, an
 appealing possibility consists of the inclusion of high-energy
(small-$x$) resummation effects in the splitting 
functions~\cite{Altarelli:2008aj,Ciafaloni:2007gf} and their matching with the 
fixed order expressions,
 a feature
which is not present in any public evolution code yet.
Another possible interesting extension, where more theory
work is also required, would be the implementation of the pure
electroweak corrections to PDF 
evolution~\cite{Ciafaloni:2001mu,Ciafaloni:2005fm}.
In order to improve the perturbative accuracy
of the QED evolution in the DGLAP evolution,  one
could attempt to achieve higher-order accuracy in $\alpha$. 
This would imply the inclusion of the mixed splitting
functions, proportional to $\mathcal{O}\lp \alpha\alpha_s\rp$,
as well as of the $\mathcal{O}\lp \alpha^2\rp$ corrections,
 requiring however substantial modifications in the structure of {\tt APFEL}.
Including higher-order corrections in the QED expansion
  might eventually be required when future, more
accurate LHC data in processes like low- and high-mass Drell-Yan production and
high invariant mass $WW$ production become available.
Another possibility worth exploring in this sense is the use 
of precise data on photo-production at HERA~\cite{Klasen:2002xb} 
to obtain additional information of the photon PDF.

\bigskip
\bigskip

The {\tt APFEL} program is available from its {\tt HepForge} website:
\begin{center}
{\bf \url{http://apfel.hepforge.org/}~}
\end{center}
and it can also be accessed directly from the svn repository, both
the development trunk:
\begin{center}
\tt svn checkout http://apfel.hepforge.org/svn/trunk apfel
\end{center}
as well as the current stable release:
\begin{center}
\tt svn checkout http://apfel.hepforge.org/svn/tags/2.0.0 apfel-2.0.0
\end{center}

\bigskip

{\bf\noindent  Acknowledgments \\}
We are grateful to Stefano Forte for discussions about
QED corrections in PDFs, to Gavin Salam for his comments on the
manuscript and to Stefan Weinzierl for providing
 useful information about the {\tt partonevolution1.1.3} package.
We also thank Emanuele Nocera for testing the {\tt GUI} and providing 
useful feedback.
We finally thank the referee for insightful comments about the differences
between the various options for solving the coupled QCD$\otimes$QED
evolution equations in {\tt APFEL}.
V.~B. is supported by the ERC grant 291377, 
"LHCtheory: Theoretical predictions and analyses of LHC physics: 
advancing the precision frontier".
S.~C. is supported by an Italian PRIN 2010 and by a European
EIBURS grant. 
J.~R. is partially supported by a Marie Curie 
Intra--European Fellowship of the European Community's 7th Framework 
Programme under contract number PIEF-GA-2010-272515.

\clearpage

\bibliography{qedevol}

\end{document}